\documentclass[journal]{IEEEtran}

\usepackage{xcolor}
\usepackage{cite}
\usepackage{amsmath,amssymb,amsfonts,mathtools}
\usepackage{algorithmic}
\usepackage{graphicx}
\usepackage{textcomp}
\usepackage[font=footnotesize]{caption}
\usepackage[font=footnotesize]{subcaption}
\usepackage{enumerate}

\newtheorem{lemma}{Lemma}
\newtheorem{remark}{Remark}
\newtheorem{assump}{Assumption}
\newtheorem{theorem}{Theorem}
\newtheorem{corollary}{Corollary}

\begin{document}

\title{2-D Directed Formation Control Based on Bipolar Coordinates} 

\author{Farhad Mehdifar, Charalampos P. Bechlioulis, Julien M. Hendrickx, and Dimos V. Dimarogonas 
\thanks{F. Mehdifar and D. V. Dimarogonas are with the Division of Decision and Control Systems, KTH Royal Institute of Technology, Stockholm, Sweden (e-mail: mehdifar@kth.se; dimos@kth.se). F. Mehdifar and D. V. Dimarogonas are supported by ERC CoG LEAFHOUND, H2020-ICT project CANOPIES, and the KAW foundation. Part of this work was performed while F. Mehdifar was at UCLouvain, under a FRIA-fellowship.}
\thanks{C. P. Bechlioulis is with the Division of Systems and Control of the Department of Electrical and Computer Engineering at University of Patras, Patra, Greece (e-mail: chmpechl@upatras.gr).}
\thanks{J. M. Hendrickx is with INMA, ICTEAM institute, UCLouvain, Louvain-la-Neuve, Belgium (e-mail: julien.hendrickx@uclouvain.be). J. M. Hendrickx is supported by the “RevealFlight” Concerted Research Action (ARC) of the Federation Wallonie-Bruxelles and “Learning from Pairwise Comparisons” incentive grant (MIS) of the F.R.S.-FNRS.}
}

\maketitle

%################################################################
\begin{abstract}
This work proposes a novel 2-D formation control scheme for acyclic triangulated directed graphs (a class of minimally acyclic persistent graphs) based on bipolar coordinates with (almost) global convergence to the desired shape. Prescribed performance control is employed to devise a decentralized control law that avoids singularities and introduces robustness against external disturbances while ensuring predefined transient and steady-state performance for the closed-loop system. Furthermore, it is shown that the proposed formation control scheme can handle formation maneuvering, scaling, and orientation specifications simultaneously. Additionally, the proposed control law is implementable in agents’ arbitrarily oriented local coordinate frames using only low-cost onboard vision sensors, which are favorable for practical applications. Finally, a formation maneuvering simulation study verifies the proposed approach.
\end{abstract}

%################################################################

%################################################################
\section{Introduction}
\label{sec:intro}
\IEEEPARstart{F}{ormation} control of multi-agent systems has been studied extensively during the past decade and depending on the sensing and controlled variables existing works can be mainly categorized into \cite{oh2015survey,anderson2008rigid}: position-based \cite{ren2007information}, displacement-based \cite{ji2007distributed}, distance-based \cite{krick2009stabilisation, yu2009control}, bearing-based \cite{zhao2015bearing}, and angle-based \cite{chan2020angle, jing2019angle, buckley2021infinitesimal, chen2020angle} methods. For more recent classes of formation control approaches as well as a comparative literature review on issues related to target formation's constraints, required measurements, and convergence, see \cite{cao2019ratio, kwon2020generalized, kwon2020hybrid, kwon2019hybrid}. Among the above categories, the position-based method requires agents to have a common knowledge of a global coordinate system while the displacement-based (also known as consensus-based) and bearing-based methods require agents' local coordinate frames to have a common orientation (i.e., to be aligned). On the other hand, \textit{coordinate-free} methods (e.g., distance-, angle-based, etc., \cite{krick2009stabilisation, yu2009control, jing2019angle, chen2020angle, buckley2021infinitesimal, cao2019ratio} are more attractive formation control architectures since they impose less implementation issues compared to other control methods. Indeed, in coordinate-free formation control, the desired shape is defined by a set of coordinate-free variables (e.g., distances, angles), which specify formation errors for agents. In addition, agents also require measurements of vectorized relative information of their neighboring agents (e.g., relative positions or bearings) in their local coordinate frames to constitute a control law. Hence, coordinate-free approaches enable us to design formation control laws in agents’ local coordinate frames, which do not require global position measurements (e.g., using GPS) nor the assumption of agents' aligned local coordinate frames (e.g., using a compass or orientation alignment methods through inter-agent communication) \cite{oh2015survey, meng2016formation, oh2013formation}.

Most of the existing results on coordinate-free formation control are developed under the assumption of bidirectional sensing (undirected sensing graph) among agents, which usually rely on different types of graph rigidity notions (e.g., distance, angle, ratio of the distances, weak,  hybrid rigidity, etc.) \cite{anderson2008rigid, cao2019ratio, chen2020angle, jing2019angle, kwon2020generalized, kwon2020hybrid}. However, it is often more practical to consider directed sensing among agents since: (i) the sensing limitations of the agents may enforce such structures, and (ii) it inherently avoids the issue of measurement mismatches in undirected formation control problems \cite{mou2015undirected,de2014controlling}. In this respect, the notion of persistent graphs was developed as the directed counterpart of distance rigidity \cite{hendrickx2008graphs,hendrickx2007directed,yu2007three}. Some earlier control designs for persistent formations include \cite{yu2009control, summers2011control, kang2015design}.

Unfortunately, most coordinate-free formation control methods only guarantee local, not global, convergence to the desired shape. Indeed, they rely on controlling the agents to satisfy certain shape constraints. But the minimal number of shape constraints may allow for multiple (but finite) shapes. In this respect, depending on the initial positions of the agents, meeting the formation constraints may not necessarily lead the agents to the correct shape. This is widely known as reflection, flip and flex ambiguities in distance-based formation control literature \cite{anderson2008rigid, oh2015survey, kwon2020hybrid}. In particular, distance rigidity theory (especially when the target formation is not globally rigid) cannot distinguish shapes under reflections, flip or flex ambiguities merely with distance constraints between agents, thus convergence to the desired shape specified only with inter-agent distances is not guaranteed. These ambiguity and local convergence issues also remain for angle, ratio of the distances, and weak rigidity notions as well \cite{jing2019angle, chen2020angle, cao2019ratio, kwon2020generalized}. To tackle these issues, in the context of undirected and directed 2-D distance-based formations some recent works have employed extra types of formation constraints (e.g., signed area and edge/signed angle) along with the inter-agent distances for characterizing the desired formation uniquely to establish (almost) global shape convergence \cite{anderson2017formation, cao2019almost,sugie2018hierarchical,sugie2020global, liu2019directed, liu2020distanceplus, kwon2019hybrid, kwon2020hybrid}. However, it turns out that imposing additional formation constraints leads to unwanted equilibria. In particular, since the distance and the auxiliary formation constraint (i.e., signed area/angle) interfere with each other at certain agent positions, new undesirable equilibria emerge, limiting the existing results to desired shapes under certain conditions. In addition, these approaches usually lead to tedious control gain tuning, which complicates controller design process and its extension to more practical formation control problems, where agents may have more complex tasks or dynamics, etc., \cite{anderson2017formation, cao2019almost,sugie2020global, sugie2018hierarchical ,liu2019directed, liu2020distanceplus}. In contrast to these works, recently \cite{liu2020distance} used orthogonal error variables for characterizing 2-D directed distance-based formations with (almost) global shape convergence. An extension of this approach for 3-D directed distance-based formations is proposed in \cite{liu2021orthogonal}. A completely different method for global convergence of directed distance-based formations is proposed in \cite{kang2016distance} that relies on calculating desired target points. 

Another advantage of coordinate-free control approaches concerns reducing agents' costs since they require less complex equipment for sensing and local interactions. Up to now, most of the coordinate-free formation control methods require relative position measurements for all agents whereas a few of them (e.g., \cite{cao2019ratio,chen2020angle}) only require bearing (or vision-based) measurements. In this respect, since bearing information is easier to obtain through onboard cameras, it is more favorable in practical applications \cite{tron2016distributed}. To the best of our knowledge, all coordinate-free formation control methods that have been developed to deal with global shape convergence also require all agents to measure the relative positions.

In practical formation control problems, agents are not only required to maintain the desired shape but also need to cooperatively move (maneuver) while obtaining certain formation orientations and scalings. Nevertheless, most results on coordinate-free formation control focus on stabilization of stationary formations \cite{oh2015survey, cao2019ratio, chen2020angle}. On the other hand, most of the existing formation maneuvering results are mainly limited by the assumption of aligned local coordinate frames of agents (or equivalently existence of inter-agent communications to share velocity-related information) and tracking constant reference velocities \cite{deghat2015combined, yang2018distributed,cai2015formation , mehdifar2018finite, kang2015design, chen2020angle2}, whereas only a few results are developed to discard these limitations \cite{mehdifar2020prescribed, kang2018shape}. Moreover, the problems of orientation and scaling control are usually handled separately \cite{sun2017distributed2,yang2019stress} and are not integrated into the maneuvering task. Recently, \cite{chen2021maneuvering} has proposed an angle-based formation maneuvering control with orientation and scaling adjustment, however, this result is applied to undirected formations with local shape convergence and (piece-wise) constant reference velocities. Moreover, \cite{li2020layered} has considered a layered affine formation maneuver problem for directed ($d+1$-rooted) graphs in which the formation scale, orientation, and also shearing can be adjusted by changing the configuration among $d+1$ leaders in $d$-dimensional space formations. However, this result still relies on the relative position and velocity measurements and requires a sufficiently large number of edges (sensing links) in the directed graph modeling inter-agent interactions, which is far from minimal.

The existence of external disturbances that affect the agents' dynamics is a significant issue of practical interest for multi-agent formation applications. It is noteworthy to mention that in coordinate-free formation control problems, only a few recent works have taken into account external disturbances and uncertainties in agents' dynamics \cite{bae2018disturbance, van2020distance, mehdifar2020prescribed}, with these results only applying to local shape convergence. Finally, another crucial issue concerns the transient response of the multi-agent formations. In this regard, Prescribed Performance Control (PPC) \cite{bechlioulis2008robust, bechlioulis2014low}, proposes a simple and constructive procedure based on which the transient performance of the closed-loop system is predetermined by certain user defined performance bounds. Recently, PPC has been utilized for displacement-based and tree structure formation control problems \cite{bechlioulis2016decentralized, verginis2019robust,chen2020leader}, as well as distance-based formation control \cite{mehdifar2020prescribed}.

In this paper, we propose a robust 2-D directed coordinate-free formation control using bipolar coordinates with (almost) global shape convergence (i.e., global convergence except for a zero-measure set of initial conditions) and guaranteed transient and steady-state performance. The target formation and the sensing topology among agents are defined by a triangulated acyclic minimally persistent graph (constructed under Assumption \ref{assu:G} in Section \ref{Sec:prob_formu}), which constitutes a (distance-rigid) directed hierarchical leader-follower structure with a minimum possible number of edges, in which: agent 1 (leader) is only responsible for formation translations, agent 2 (secondary-leader) only follows agent 1 and is responsible for formation scaling and orientation adjustments, and the rest of the agents (followers), where each of them follows exactly two other agents, are responsible for generating and maintaining the desired shape. In particular, given a desired formation, first, we show that the desired position of each follower agent can be uniquely characterized with respect to a local bipolar coordinate system assigned to it, with its two neighbors as the foci of the bipolar coordinate system \cite{happel2012low, weisstein2002crc}. As a result, this leads to having a unique pair of desired bipolar coordinates values, i.e., a desired angle and a desired (logarithm of) ratio of distances w.r.t. the two foci (neighbors), which characterize each follower's formation errors (see Fig.\ref{fig:local_cartes_bipolar} for illustrations). Then, leveraging the fact that each follower's formation errors can be reduced (independently) by moving along the two orthogonal directions of its associated bipolar coordinate basis, we employ the prescribed performance control methodology to design robust controllers under external disturbances stabilizing the formation errors, thus achieving (almost) global convergence. In the control design procedure, user-defined performance guarantees on the system's response are achieved by imposing time-varying decreasing performance bounds (constraints) on the formation errors. It is also important to note that, keeping the formation errors within some desirable decreasing performance bounds not only introduces robustness w.r.t. external disturbance on agents' motion dynamics, but also helps us to design the formation controllers of agent 2 and the followers to handle formation maneuvering (in the presence of a moving leader) with time-varying scale and orientation adjustments. Moreover, owing to the usage of local bipolar coordinates for characterizing the formation errors, instead of relative position measurements, the proposed control laws only require bearing and ratio of the distances measurements for the followers, which can be obtained through onboard vision sensing (see Remark \ref{re:vision_sense} in Section \ref{sec:fmerr} for more details). 

The contributions of this work are summarized as follows:

\begin{itemize}
	\item For the first time, bipolar coordinates are employed for solving 2-D coordinate-free formation control problems with (almost) global shape convergence without introducing undesired equilibria and gain tuning issues existing in many previous works, e.g., \cite{anderson2017formation, cao2019almost,sugie2018hierarchical,sugie2020global, liu2019directed, liu2020distanceplus}.
	
	\item In contrast to the existing coordinate-free formation control results with (almost) global shape convergence \cite{anderson2017formation, cao2019almost,sugie2018hierarchical,sugie2020global, liu2019directed, liu2020distanceplus,  kwon2020hybrid, kang2016distance,liu2020distance}, our approach can handle coordinate-free formation maneuvering (with time-varying reference velocity) along with scaling and orientation adjustments. Moreover, while the above-mentioned results require all agents to measure relative positions, our approach builds upon bearing and ratio of the distances measurements that are readily available by vision (camera) sensors and thus easier to obtain in practical applications. To the best of the authors' knowledge, this work is the first one to provide such results for (almost) globally converging coordinate-free formation control. 
	
	\item To the best of our knowledge, there are no previous works on (almost) globally converging coordinate-free formations with guaranteed performance and robustness with respect to external disturbances/dynamical uncertainties.
	
	\item In contrast to relevant undirected angle-based formation control results with local convergence, e.g., \cite{chen2021maneuvering, buckley2021infinitesimal}, our approach handles robust directed formation control with (almost) global shape convergence. Moreover, the method in \cite{buckley2021infinitesimal} requires explicit communication among neighboring agents, while ours is communication-free. 
\end{itemize}

%################################################################
\section{Problem Formulation}
\label{Sec:prob_formu}
Consider a multi-agent system comprised of $n$ mobile robots on a 2-D plane governed by the following dynamics:
\begin{equation}
	\label{eq:singledyn}
	\dot{p}_i=u_i + \delta_i(t), \quad i=1,\ldots,n,
\end{equation}
where $p_i \in \mathbb{R}^2$ and $u_i \in \mathbb{R}^2$ are the position and the velocity control input of agent $i$ expressed with respect to a global coordinate frame, respectively. Let $\delta_i(t) \in \mathbb{R}^2$ represent an unknown, bounded and piece-wise continuous external disturbance on agent $i$ (e,g., wind gusts), which may also account for model uncertainties. Notice that the upper bound of the disturbances is not known a priori. 

Let the sensing topology among agents be modeled by a directed graph $\mathcal{G} = (\mathcal{V},\mathcal{E})$, where $\mathcal{V}=\{1,2,\ldots,n\}$ is the set of vertices representing the agents and $\mathcal{E}\subseteq \{(j,i)\lvert j,i \in \mathcal{V}, j \neq i\}$ such that if $(j,i) \in \mathcal{E} \Rightarrow (i,j) \notin \mathcal{E}$ is the set of directed edges depicting the directed sensing relations among the agents. More precisely, $(j,i) \in \mathcal{E}$ denotes an edge that starts from vertex $j$ (source) and sinks at vertex $i$, and its direction is indicated by $j \rightarrow i$. For $(j,i)$ we say $i$ is the \textit{neighbor} of $j$. The relative position vector corresponding to the directed edge $(j,i)$ is defined as:
\begin{equation} \label{eq: rel_pos}
	{p}_{ji} = p_i - p_j, \quad (j,i) \in \mathcal{E},
\end{equation}
and its associated \textit{relative bearing} vector $z_{ji} \in \mathbb{R}^2$ is: 
\begin{equation} \label{eq:bearing}
	z_{ji} = \dfrac{{p}_{ji}}{\| {p}_{ji} \|}, \quad (j,i) \in \mathcal{E}.
\end{equation}
In particular, in this paper the physical meaning of the directed edge $(j,i) \in \mathcal{E}$ is that only agent $j$ can measure the \textit{relative bearing} of agent $i$ with respect to itself, i.e., $z_{ji}$, and not vice versa. As will be highlighted in the sequel, we will additionally assume that for the particular case of agent $2$, which we will call as \textit{secondary leader}, not only can measure the relative bearing of agent 1, but also the absolute distance from it.

We also assume that the graph $\mathcal{G}$ is triangulated and imposes a hierarchical structure, where agent 1 is the \textit{leader}, agent 2 is the \textit{secondary leader} with agent 1 acting as its only neighbour, and agents $i \geq 3$ are the \textit{followers} with each one having exactly two neighbors to follow with smaller indices. Hence, we impose the following assumption for constructing $\mathcal{G}$:
\begin{assump} \label{assu:G}
	The directed sensing graph $\mathcal{G}$ is constructed such that:
	\begin{enumerate}
		\item $\mathrm{out}(1) = 0$, $\mathrm{out}(2) = 1$, and $\mathrm{out}(i) = 2$, $\forall i \geq 3$;
		\item If there is an edge between agents $i$ and $j$, where $i < j$, the direction must be $j \rightarrow i$;
		\item If $(k,i), (k,j) \in \mathcal{E}$ then $(j,i) \in \mathcal{E}$,
	\end{enumerate}
\end{assump}
where $\mathrm{out}(i)$ denotes the out-degree of vertex $i$ that is the number of edges in $\mathcal{E}$ whose source is vertex $i$ and whose sinks are in $\mathcal{V} \setminus \{i\}$. 

\begin{figure}[!tb]
	\centering
	%	\flushleft
	\begin{subfigure}[t]{0.24\textwidth}
		%		\centering
		\centering
		\includegraphics[width=\textwidth]{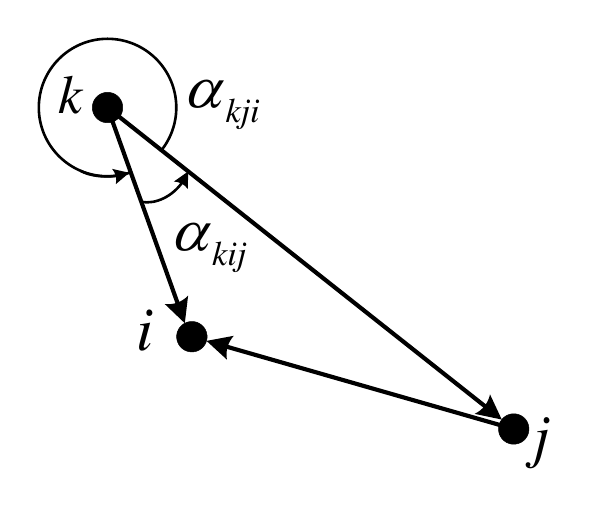}
		\caption{}
		\label{fig:trig_framework}
	\end{subfigure}%\hspace{2cm}
	~
	\begin{subfigure}[t]{0.23\textwidth}
		%		\centering
		\centering
		\includegraphics[width=\textwidth]{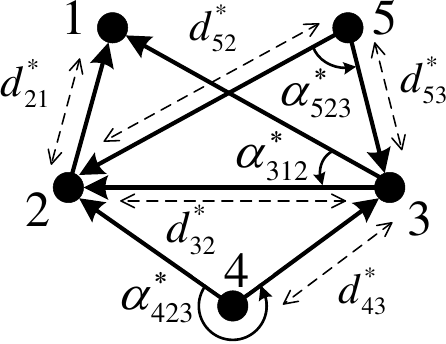}
		\caption{}
		\label{fig:angle_example}
	\end{subfigure}%\hspace{2cm}
	\caption{(a) edge-angle in a triangular subgraph. (b) example of a desired formation (note that $d_{31}^{\ast} = d_{52}^{\ast}$ and $d_{42}^{\ast} = d_{43}^{\ast}$).}
	\label{fig:edg_angle}
\end{figure}

We highlight that cases 1) and 2) in Assumption \ref{assu:G} impose $\mathcal{G}$ to be acyclic\footnote{Notice that a directed path in a graph is a sequence of vertices of finite length such that from each of its vertices there is a directed edge to the next vertex in the sequence. A directed graph is acyclic if there is no directed path in the graph that starts and ends with the same node.} minimally persistent with edge set cardinality $|\mathcal{E}|=2n-3$ \cite{hendrickx2007directed, liu2020distanceplus} (for more information see Remarks \ref{rem:rigidity} and \ref{rem: minimal} and references therein), while case 3) establishes triangulation in $\mathcal{G}$. Note that under Assumption \ref{assu:G}, $\mathcal{G}$ is composed of acyclic directed triangles (i.e., triangular sub-graphs as depicted in Fig.\ref{fig:trig_framework}). Fig.\ref{fig:angle_example} shows an example of $\mathcal{G}$ constructed under Assumption \ref{assu:G}. 

For each follower $k$ in $\mathcal{G}$ with two neighbors $i$ and $j$, we can define an \textit{edge-angle} as the angle $\alpha_{kij} \in [0, 2 \pi)$ formed by the edges $(k,i), (k,j) \in \mathcal{E}$, measured by convention counterclockwise from edge $(k,i)$ to edge $(k,j)$ \cite{liu2020distanceplus}. Fig.\ref{fig:trig_framework} shows the edge-angle $\alpha_{kij}$ assigned to the $k$-th follower in a (acyclic) directed triangular sub-graph of $\mathcal{G}$, where Assumption \ref{assu:G} establishes the ordering of $i<j<k$ as well. 
Based on the bearing vectors $z_{ki}$ and $z_{kj}$, the edge-angle $\alpha_{kij}$ can be obtained by:
\begin{equation}\label{eq:edge_angle_bear}
	\alpha_{kij} =
	\begin{cases}
			\arccos(z_{ki}^T z_{kj}) & \text{if} \, (z_{ki}^{\bot})^T z_{kj} \geq 0, \\
			2\pi - \arccos(z_{ki}^T z_{kj}) & \text{otherwise,}
		\end{cases}
\end{equation}
in which $z_{ki}^{\bot} \coloneqq J z_{ki}$, where
\begin{equation} \label{eq:rotation90counter}
	J  =
	\begin{bmatrix*}[r]
			0 & -1 \\
			1 & 0
		\end{bmatrix*}
\end{equation}
denotes the $\frac{\pi}{2}$-counterclockwise rotation matrix. 

It is known that based on the directed sensing graph $\mathcal{G}$ (respecting Assumption \ref{assu:G}), we can \textit{uniquely} define a desired formation characterized by \cite{liu2020distanceplus}: 
\begin{enumerate}[(i)]
	\item A set of $2n-3$ \textit{desired distances} $d_{ji}^{\ast}$, appointed to the directed edges $(j,i) \in \mathcal{E}$.
	\item A set of $n-2$ \textit{desired edge-angles} $\alpha_{kij}^{\ast}, (k,i), (k,j) \in \mathcal{E} \setminus \{(2,1)\}$, $i<j<k$, (see Fig.\ref{fig:angle_example} for an example). 
\end{enumerate}

Given a desired formation characterized by a graph $\mathcal{G}$ (under Assumption 1) and the corresponding sets of desired distances and edge-angles, the objective is to design a decentralized robust control protocol for \eqref{eq:singledyn} such that:
\begin{subequations} \label{eq:objective}
	\begin{alignat}{3}
			\| p_j(t)  - p_i(t) \| &\rightarrow d_{ji}^{\ast} \quad &&\mathrm{as} \quad t \rightarrow \infty, \label{eq:distance_to_desired} \\
			\alpha_{kij}(t) &\rightarrow \alpha_{kij}^\ast \quad &&\mathrm{as} \quad t \rightarrow \infty, \label{eq:angle_to_desired}
	\end{alignat}
\end{subequations}
for all $(j,i) \in \mathcal{E}$ and $(k,i), (k,j) \in \mathcal{E} \setminus \{(2,1)\}$, $i<j<k$, respectively, while avoiding zero distance among neighboring agents (i.e., $\|{p}_{ji}\|\neq 0, \forall(j,i) \in \mathcal{E}, \forall t\geq 0$) so that all edge-angles are well-defined. It is known that, satisfaction of \eqref{eq:objective} is equivalent to strong congruency \cite{liu2019directed,liu2020distanceplus} between the actual formation of the agents and the desired formation (see Lemma 1 of \cite{liu2020distanceplus}). This means that if \eqref{eq:objective} gets satisfied, the agents can achieve the desired formation only up to rotations and translations \cite{liu2020distanceplus}.  

\begin{remark} \label{rem:rigidity}
 Assumption \ref{assu:G} indicates that $\mathcal{G}$ is a leader-first-follower type formation \cite{summers2011control}, which belongs to a class of acyclic minimally persistent graphs \cite{summers2011control, hendrickx2007directed}. Persistent graphs are the directed counterpart of undirected distance rigid graphs \cite{anderson2008rigid,hendrickx2008graphs}. Rigid graphs and rigidity theory has been widely used as a tool for studying coordinate-free formations \cite{anderson2008rigid, krick2009stabilisation, cao2019ratio, chen2020angle}. In particular, distance rigidity (persistence) of an undirected (directed) graph ensures that the desired formation can be characterized merely by a set of desired distances. However, in general, such formation characterization is not unique and suffers from local shape convergence and reflection issues due to the existence of undesired shapes (known as reflections, flip and flex ambiguities) satisfying the given set of desired distances (see \cite{anderson2008rigid,anderson2017formation, liu2019directed} for examples). To tackle these issues, extra types of formation parameters (e.g., signed area, edge-angle, etc.) have been recently employed along with the distances to characterize the desired formation uniquely, which is necessary for having global shape convergence \cite{anderson2017formation,liu2019directed,cao2019almost,sugie2018hierarchical,sugie2020global,liu2020distanceplus}. More precisely, the notion of strong congruency defined in \cite{liu2019directed,liu2020distanceplus} is related to distinguishing the shape of congruent rigid frameworks (please refer to \cite{anderson2008rigid} for a definition) and thus avoiding reflected frameworks, see \cite[Section II.c]{liu2020distanceplus}. In other words, in general, congruent rigid frameworks may have the issue of position reflections and are not shape-preserving, while strong congruency removes this issue by exploiting an additional parameter (i.e., edge-angle or signed area) to characterize the position of each vertex in a rigid framework. As an example, consider distances $\|p_{ki}\|, \|p_{kj}\|$ and the edge-angle $\alpha_{kij}$ in Fig.\ref{fig:trig_framework}, where $0 < \alpha_{kij} < \pi$. If the position of vertex $k$ is reflected without altering its distances with respect to $i$ and $j$, then $\pi < \alpha_{kij} < 2\pi$. This property will allow us to distinguish the position of agent $k$ from its reflection with respect to the line passing through agents $i$ and $j$. This is further depicted in Fig.\ref{fig:angle_example} by comparing the desired edge-angles assigned to agents 4 and 5. 
\end{remark}
\begin{remark} \label{rem: minimal}
	 From Assumption \ref{assu:G}, $\mathcal{G}$ is minimally persistent, meaning that it is a persistent graph with minimum number of edges. This is favorable in practice since it requires minimum number of relations (sensing) among agents. This assumption on $\mathcal{G}$ is not restrictive since $\mathcal{G}$ can be easily adapted to any geometrical shape and scaled up to any number of agents through the Henneberg type I construction \cite{bereg2005certifying,liu2019directed}. More precisely, $\mathcal{G}$ in our paper is also referred to as a \textit{directed triangulated Laman graph} (see \cite{babazadeh2020distance} for more details). 
\end{remark}

%################################################################
\section{Main Results}
\label{sec:main}
In this section, we will first introduce two independent variables based on bipolar coordinates that characterize the desired positions of the followers within the desired formation. Then, leveraging the fact that each follower's formation errors can be reduced (independently) by moving along the two orthogonal directions of its associated bipolar coordinate basis, we use the prescribed performance control (PPC) method to design proper decentralized robust formation controllers to meet \eqref{eq:objective} in the presence of external disturbances with (almost) global convergence to the desired shape. In particular, for the controller design procedure, first, a desirable transient and steady-state performance is imposed by using time-varying decreasing performance bounds on the formation errors and it is shown that by fine-tuning the performance bounds one can also ensure non-collocation of the neighboring agents so that the edge-angles always remain well-defined. Then, a nonlinear transformation is used to map the constrained error to an unconstrained one whose stability merely ensures the satisfaction of proposed time-varying error constraints. At the end of this section, the main theorems and stability analysis are provided along with an extension of agent 2's controller for adjusting formation orientation.
\subsection{Characterization of the Desired Formation Based on Bipolar Coordinates}
\begin{figure}[tb]
	\centering
	%	\flushleft
	\begin{subfigure}[t]{0.40\textwidth}
		\centering
		\includegraphics[width=\textwidth]{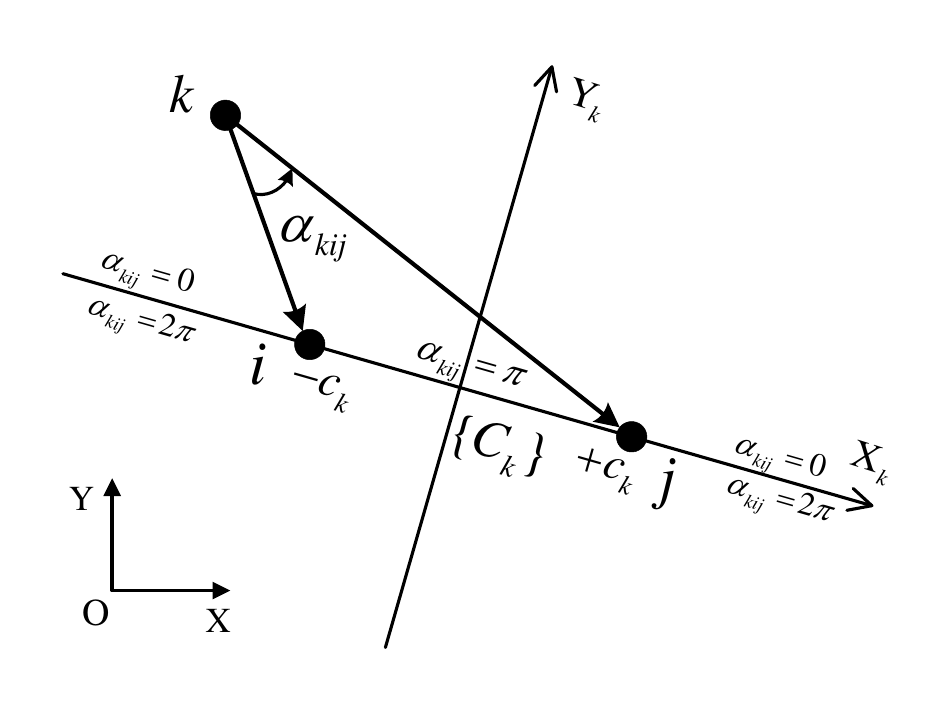}
		\caption{}
		\label{fig:local_cart}
	\end{subfigure}
	
	\begin{subfigure}[t]{0.45\textwidth}
		\centering
		\includegraphics[width=\textwidth]{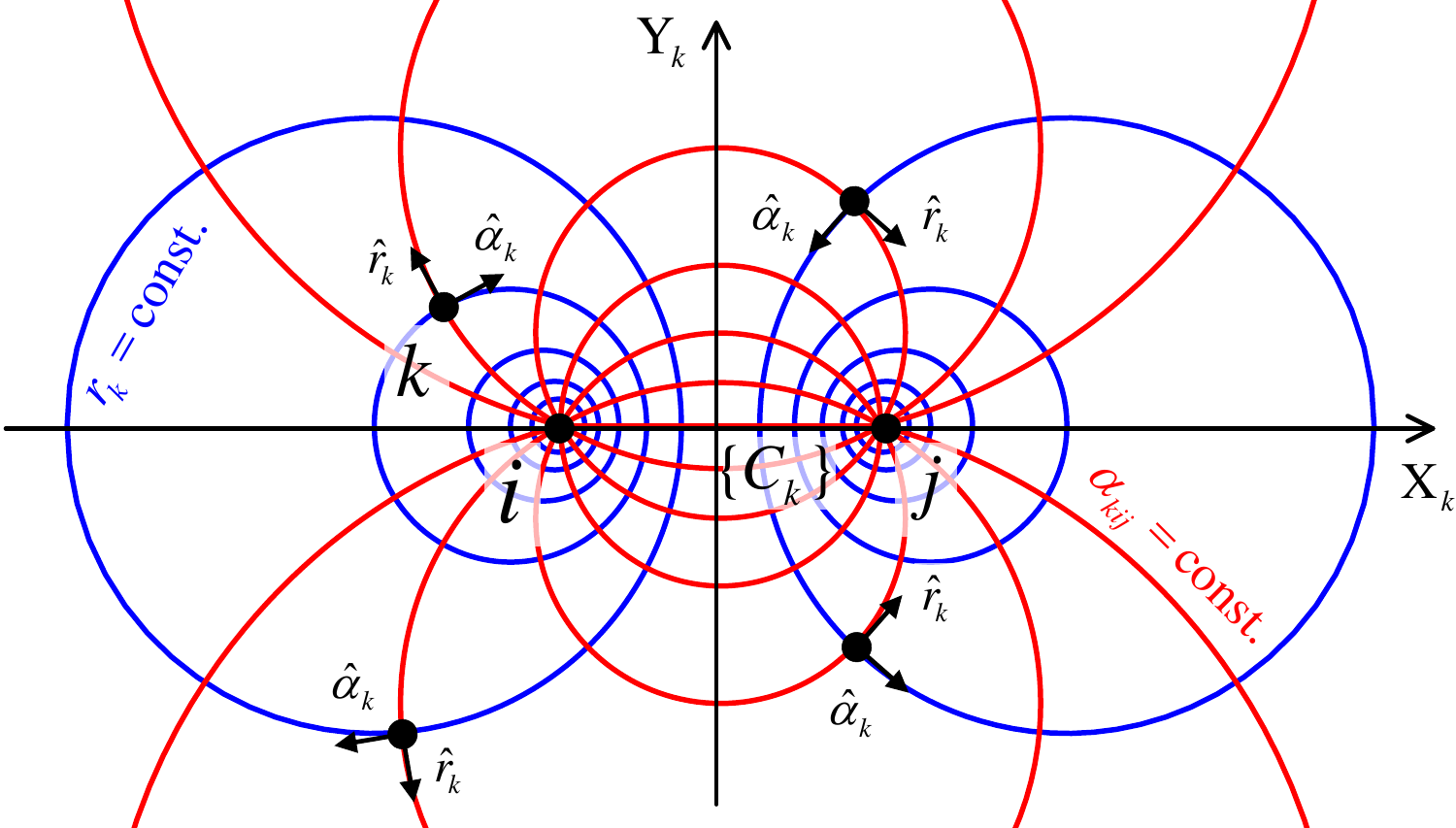}
		\caption{}
		\label{fig:local_bipolar}
	\end{subfigure}
	\caption{(a) The virtual local Cartesian coordinate frame $\left\lbrace C_k \right\rbrace$ uniquely characterizes the position of agent $k\geq3$ with respect to its neighbors (agents $i$ and $j$). Instead of using the Cartesian coordinates in $\left\lbrace C_k \right\rbrace$ one can adopt bipolar coordinates \eqref{eq:edge_angle_bear} and \eqref{eq:log_ratio} in $\left\lbrace C_k \right\rbrace$ to determine agent $k$'s position. (b) Orthogonal bipolar coordinate basis $\widehat{r}_k$, $\widehat{\alpha}_k$ associated with agent $k\geq3$ and some of their isoquant curves.}
	\label{fig:local_cartes_bipolar}
\end{figure}

Consider a triangular sub-graph of $\mathcal{G}$ as in Fig.\ref{fig:trig_framework} where $i<j<k$. If $\|{p}_{ji}\|\neq 0$ one can define a virtual local Cartesian coordinate frame based on vertices $i$ and $j$, denoted by $\left\lbrace C_k \right\rbrace$ as in Fig.\ \ref{fig:local_cart}, with its origin located in the middle of the $i$-$j$ line segment. Note that the position of node $k$ can be uniquely determined in $\left\lbrace C_k \right\rbrace$ w.r.t. its neighboring agents $i$ and $j$. It is also known that agent $k$'s position in $\left\lbrace C_k \right\rbrace$ can be expressed by the bipolar coordinates $(r_k, \alpha_{kij}) \in \mathbb{R}^2$ associated with $\left\lbrace C_k \right\rbrace$, where nodes $i$ and $j$ are the two foci of the bipolar coordinate system \cite{happel2012low}. 
Recall that the bipolar coordinate variable $\alpha_{kij}$ (edge-angle) was already introduced in Section \ref{Sec:prob_formu} and it is given by \eqref{eq:edge_angle_bear}. Moreover, $r_k$ is the natural logarithm of the \textit{ratio of the distances} $r_{kij} \coloneqq \| {p}_{ki} \| / \| {p}_{kj} \|, (k,i), (k,j) \in \mathcal{E} \setminus \{(2,1)\}$, $i<j<k$, between node $k$ and the foci $i$ and $j$, expressed by:
\begin{equation}\label{eq:log_ratio}
	r_k \coloneqq \ln r_{kij} = \ln \dfrac{\| {p}_{ki} \|}{\| {p}_{kj} \|}, \quad (k,i), (k,j) \in \mathcal{E} \setminus \{(2,1)\}, 
\end{equation}
where $r_k \in \mathbb{R}$. Note that, when agent $k$ approaches one of the foci $i$ or $j$ (i.e., either $\| {p}_{ki} \|\rightarrow 0$ or $\| {p}_{ki} \| \rightarrow 0$), $r_k$ tends to $\pm \infty$. The bipolar coordinates are related to the $\left\lbrace C_k \right\rbrace$ frame with the following (almost) one-to-one (except at the foci of the bipolar coordinates, $i$ and $j$) transformation \cite{weisstein2002crc}:
\begin{subequations}\label{eq:bipolar_cord_in_virtual}
	\begin{align}
		x_k^{[C_k]} &= c_k  \dfrac{\sinh r_k}{\cosh r_k - \cos \alpha_{kij}}, \\
		y_k^{[C_k]} &= c_k  \dfrac{\sin \alpha_{kij}}{\cosh r_k - \cos \alpha_{kij}},
	\end{align}
\end{subequations}
where $p_k^{[C_k]} = [x_k^{[C_k]}, y_k^{[C_k]}]^T \in \mathbb{R}^2$ is the position of vertex $k$ with respect to frame $\left\lbrace C_k \right\rbrace$ and $c_k = 0.5  \| {p}_{ji}\| > 0, k\geq3$.

The bipolar coordinate system $(r_k, \alpha_{kij})$ is indeed a 2-D orthogonal curvilinear coordinate system \cite{happel2012low,weisstein2002crc} (similar to the well-known polar coordinate system), therefore, one can define a local orthogonal basis at each point in the 2-D plane of $\left\lbrace C_k \right\rbrace$ showing the directions of increase for $\alpha_{kij}$ and $r_k$. Fig.\ \ref{fig:local_bipolar} shows orthogonal bipolar coordinates basis $\widehat{\alpha}_{k} \in \mathbb{R}^2$ and $\widehat{r}_k \in \mathbb{R}^2$ associated with $\left\lbrace C_k \right\rbrace$ at some arbitrary points of interest as well as some $\alpha_{kij}$ and $r_k$ isoquant curves that create circles centered along the $\mathrm{Y}_k$ and $\mathrm{X}_k$ axis, respectively. 

Given a target formation expressed by the graph $\mathcal{G}$ along with the desired edge-angles (i.e., $\alpha_{kij}^{\ast}$) and distances (i.e., $d_{ji}^{\ast}$), we can use the desired bipolar coordinates $(r_k^{\ast}, \alpha_{kij}^{\ast}) \in \mathbb{R}^2$  to uniquely determine the desired position of agent $k\geq3$ with respect to its two neighbors $i$ and $j$ ($i<j<k$), where
\begin{equation} \label{eq:desired_ratio}
	r_k^{\ast} \coloneqq  \ln \dfrac{d_{ki}^{\ast}}{d_{kj}^{\ast}}, \quad (k,i), (k,j) \in \mathcal{E} \setminus \{(2,1)\}.
\end{equation} 
In this regard, we propose the following lemma:
\begin{lemma} \label{lem:equvalence}
	Given a desired formation shape based on a specific directed sensing graph $\mathcal{G} = (\mathcal{V}, \mathcal{E})$ under Assumption \ref{assu:G}, as well as $\alpha_{kij}^{\ast}$, $(k,i), (k,j) \in \mathcal{E} \setminus \{(2,1)\}$, $i<j<k$ and $d_{ji}^{\ast}$, $(j,i) \in \mathcal{E}$, satisfying:
	\begin{subequations}\label{eq:lemma_bipol}
		\begin{alignat}{4}
			\| p_2(t)  - p_1(t) \| &\rightarrow d_{21}^{\ast},&& \quad &&&\mathrm{as} \quad t \rightarrow \infty, \label{eq:2nd_agent_dist} \\
			r_k(t) &\rightarrow r_k^{\ast}, \quad &&k\geq 3, \quad &&&\mathrm{as} \quad t \rightarrow \infty, \label{eq:ratio_cond} \\
			\alpha_{kij}(t) &\rightarrow \alpha_{kij}^\ast,\quad &&k\geq 3, \quad &&&\mathrm{as} \quad t \rightarrow \infty, \label{eq:angle_to_desired_lem}
		\end{alignat}
	\end{subequations}
	 is equivalent to the satisfaction of \eqref{eq:objective}.
\end{lemma}
\begin{IEEEproof}
	 \eqref{eq:lemma_bipol} $\Rightarrow$ \eqref{eq:objective}: Recall that due to Assumption \ref{assu:G}, $\mathcal{G}$ is comprised of triangular sub-graphs. Every acyclic directed triangular sub graph of $\mathcal{G}$ with vertices $i, j$, and $k$  defines a triangle denoted by $\triangle_{ijk}$. Now consider the triangle (composed of agents 1, 2 and 3) of the desired formation and the actual triangle formed by the agents at time instance $t\geq0$, which are denoted by $\triangle_{123}^{\ast}$ and $\triangle_{123}(t)$, respectively. Note that, owing to the side-angle-side similarity theorem between two triangles, if the ratio of two sides as well as the angle included between these sides become identical in $\triangle_{123}^{\ast}$ and $\triangle_{123}(t)$, then we can infer that $\triangle_{123}(t)$ is similar to $\triangle_{123}^{\ast}$. Therefore, satisfaction of \eqref{eq:ratio_cond} and \eqref{eq:angle_to_desired_lem} for agent 3 (i.e., $r_3 \rightarrow r_3^{\ast}$ and $\alpha_{312} \rightarrow \alpha_{312}^{\ast}$) ensures that $\triangle_{123}(t)$ becomes similar to $\triangle_{123}^{\ast}$ in the limit. In addition, satisfaction of  \eqref{eq:2nd_agent_dist} further ensures that $\triangle_{123}(t)$ and $\triangle_{123}^{\ast}$ will have the same edge lengths in the limit, that is $\| p_3(t)  - p_1(t) \| \rightarrow d_{31}^{\ast}$ and $\| p_3(t)  - p_2(t) \| \rightarrow d_{32}^{\ast}$ as $t\rightarrow\infty$. Repeating these arguments for the rest of the triangular sub-graphs of $\mathcal{G}$ in the desired and the actual formations, i.e., $\triangle_{ijk}^{\ast}$ and $\triangle_{ijk}(t)$, $(k,i), (k,j) \in \mathcal{E} \setminus \{(2,1)\}$, $i<j<k$, will result in satisfaction of \eqref{eq:objective} for all triangular sub-graphs of $\mathcal{G}$. Therefore, \eqref{eq:lemma_bipol} implies \eqref{eq:objective}.

	\eqref{eq:objective} $\Rightarrow$ \eqref{eq:lemma_bipol}: Again consider $\triangle_{123}(t)$ and $\triangle_{123}^{\ast}$. If \eqref{eq:objective} is satisfied for $\triangle_{123}(t)$, i.e., $\| p_3(t)  - p_1(t) \| \rightarrow d_{31}^{\ast}$, $\| p_3(t)  - p_2(t) \| \rightarrow d_{32}^{\ast}$, $\| p_2(t)  - p_1(t) \| \rightarrow d_{21}^{\ast}$, and $\alpha_{312} \rightarrow \alpha_{312}^{\ast}$, then satisfaction of \eqref{eq:lemma_bipol} for $\triangle_{123}(t)$ can be readily deduced. Similarly to the previous case, by repeating these arguments for the rest of the triangular sub-graphs in the desired and the actual formations, one can  infer that, in general, \eqref{eq:objective} implies \eqref{eq:lemma_bipol} and this completes the proof. 
\end{IEEEproof}

Recall that equations \eqref{eq:distance_to_desired} and \eqref{eq:angle_to_desired} indicate that the secondary leader (i.e., agent 2) is only required to keep a certain distance with respect to the leader (agent 1), whereas the rest of the agents (i.e., followers) are required to keep a certain edge-angle and two specific distances with respect to their neighbors. Therefore, a direct approach to achieve \eqref{eq:objective} for each follower agent is to control three variables: two distances and an edge-angle \cite{liu2020distanceplus}. Alternatively, followers can use the signed area information (instead of the edge-angle) along with the distances to achieve the same objective \cite{liu2019directed,sugie2018hierarchical,cao2019almost,sugie2020global}. However, using an extra shape constraint (i.e., signed area  or edge-angle) for the followers to achieve the desired formation may introduce new undesirable equilibria as the distance and signed-area/edge-angle constraints interfere with each other at certain agent positions (see \cite{anderson2017formation,liu2019directed,sugie2018hierarchical,cao2019almost,sugie2020global,liu2020distanceplus} for more details and examples). Indeed, these variables do not always constitute an orthogonal space, in which each formation variable can be adjusted independently by moving along orthogonal directions. Lemma \ref{lem:equvalence} overcomes this issue as it only requires the followers to control only two orthogonal (i.e., independent) formation variables \eqref{eq:ratio_cond} and \eqref{eq:angle_to_desired_lem}. In the sequel, we will leverage this fact to design the formation controllers of the follower agents (see Subsection \ref{subss}), which allows for (almost) global convergence to the desired shape.

The proof of Lemma \ref{lem:equvalence} also reveals that by modifying the distance of agent 2 with respect to agent 1 (i.e., $\| p_2(t)  - p_1(t) \|$) one can change the \textit{scale} of the actual formation at the steady-state (formation scaling means maintaining all angles in the shape and increasing or decreasing all edge lengths with the same proportion). Therefore, if the secondary leader alters its desired distance with respect to the leader, by considering a time-varying desired distance $d_{21}^{\ast}(t)$, then it can control the formation's scale, which is of high importance in practical formation control applications (e.g., passing through narrow passages, obstacle avoidance, etc.). 
\subsection{Formation Errors} \label{sec:fmerr}
To quantify the control objective we define 3 types of error variables. First, the \textit{squared distance error} between agents 2 and 1 is defined as:
\begin{equation} \label{eq:eij}
	e_{d} =\|{p}_{21}\|^2-(d_{21}^{\ast}(t))^2,
\end{equation}
where $d_{21}^{\ast}(t): \mathbb{R} \rightarrow \mathbb{R}_{>0}$ is a strictly positive and continuously differentiable function of time with a bounded derivative representing the desired (in general, time-varying) distance between agents 2 and 1. Notice that $\|{p}_{21}\|=d_{21}^{\ast}(t)$ if and only if $e_d = 0$. Secondly, the \textit{logarithmic ratio of the distances error} is defined as:
\begin{equation} \label{eq:r_error}
	{e}_{r_k} = r_{k} - r_{k}^{\ast}, \quad k=3,\dots,n,
\end{equation}
where $r_{k}$ and $r_{k}^{\ast}$ are defined in \eqref{eq:log_ratio} and \eqref{eq:desired_ratio}, respectively. Third, the \textit{edge-angle error} is defined as:
\begin{equation} \label{eq:alpha_error}
	e_{\alpha_k} = \alpha_{kij} - \alpha_{kij}^{\ast}, \quad (k,i), (k,j) \in \mathcal{E} \setminus \{(2,1)\},
\end{equation}
where $\alpha_{kij}$ is defined in \eqref{eq:edge_angle_bear} and $i<j<k$ \footnote{We use the subscript $e_{\alpha_k}$ instead of $e_{\alpha_{kij}}$ for better readability.}. Note that, \eqref{eq:r_error} and \eqref{eq:alpha_error} are independent (orthogonal) error variables defined only for the followers. More precisely, by moving along each bipolar coordinates basis, $\widehat{r}_k$ and $\widehat{\alpha}_k$, each follower can reduce \eqref{eq:r_error} and \eqref{eq:alpha_error}, respectively, without affecting the other error variable. Finally, due to the above discussion and Lemma \ref{lem:equvalence}, by adopting the bipolar coordinates approach, the control objective of \eqref{eq:objective} is met by zero stabilization of the errors defined in \eqref{eq:eij}, \eqref{eq:r_error}, and \eqref{eq:alpha_error} while maintaining $\|{p}_{ji}(t)\|\neq 0, \forall (j,i) \in \mathcal{E}, \forall t\geq 0$.

\begin{remark} \label{re:vision_sense}
In order to meet \eqref{eq:ratio_cond} and \eqref{eq:angle_to_desired_lem} in Lemma \ref{lem:equvalence}, each follower is only required to sense and adjust its edge-angle formed by its neighbors, as well as the ratio of the distances with respect to them. It is known that, in general, onboard vision-based sensors (e.g., monocular cameras) give projective measurements that do not contain distance information. As a consequence, it is possible to obtain only bearing (direction) information, from which the angle information can be then retrieved \cite{tron2016distributed} (e.g., by \eqref{eq:edge_angle_bear}). Moreover, as explained in \cite[Section II.D]{cao2019ratio}, the ratio of the distances can also be extracted from a single image of a camera by comparing projections of two identical (yet unknown) sized (spherical or circular) objects/markers (i.e., two neighbors of a certain follower agent) on the image plane of a camera. In the absence of spherical (circular) shaped agents/markers, each robot may use a database of CAD models for obtaining the ratio of the distances \cite{florczyk2006robot}. Therefore, all followers are required to be equipped only with low-cost vision sensors to perceive the required information/feedback. This is in contrast to many related results in coordinate-free formation control with (almost) global shape convergence, where relative position measurements for all agents are assumed \cite{anderson2017formation, cao2019almost,sugie2018hierarchical,sugie2020global, liu2019directed, liu2020distanceplus, liu2021orthogonal, liu2020distance,kang2016distance}.
\end{remark}

%################################################################
\subsection{Controller Design}
\label{Sec:Ctrl_Design_Analysis}
In this paper, we will adopt the Prescribed Performance Control (PPC) method \cite{bechlioulis2008robust} for designing the formation control laws in order to: i) introduce robustness against external disturbances (which also allows us to deal with the formation maneuvering problem), ii) achieve predefined transient and steady-state response for each formation error $e_{h}, h \in \{ d, r_k, \alpha_k  \}, k=\{3,\cdots , n\}$, and iii) avoid singularities in the edge-angle definition when $\|{p}_{ji}\| \rightarrow 0$, for a pair $(j,i) \in \mathcal{E}$ or when either $\alpha_{kij} = 0$ or $\alpha_{kij} = 2\pi$.

Prescribed performance is achieved when the formation errors $e_{h}(t), h \in \{ d, r_k, \alpha_k  \}$, with $k=\{3,\cdots , n\}$ evolve strictly within the predefined regions that are bounded by absolutely decaying functions of time, called \textit{performance functions} \cite{bechlioulis2008robust,bechlioulis2014low}. The mathematical expression of prescribed performance is formulated by the following inequalities:
\begin{equation}\label{eq:e_h_bound}
	\begin{gathered}
		-\underline{b}_h \rho_{h}(t)  < e_h(t) < \bar{b}_{h} \rho_{h}(t), \\
		h \in \{ d, r_k, \alpha_k  \}, k=\{3,\ldots,n\},
	\end{gathered}
\end{equation}
where  $\underline{b}_h, \bar{b}_h > 0$ are arbitrary positive scaling parameters selected properly to avoid singularities in the control problem, as presented in the sequel. Moreover, $\rho_h(t): [0,\infty) \rightarrow \mathbb{R}_{\geq0}$ are user-defined continuously decaying performance functions with bounded derivatives and strictly positive limit as $t \rightarrow \infty$ (i.e., $\lim_{t \rightarrow \infty} \rho_h(t)>0$). In this work, we will adopt the following performance functions:
\begin{equation} \label{pf_rho}
	\begin{gathered}
	\rho_h(t)=\left( 1-\rho_{\infty,h}\right)  \exp(-l_h t)+ \rho_{\infty,h}, \\
	h \in \{ d, r_k, \alpha_k  \}, k=\{3,\ldots,n\}, 
	\end{gathered}
\end{equation}
where parameters $l_{h}, \rho_{\infty,h} >0$ characterize the desired transient and steady-state performance specifications on $e_h(t)$, respectively. In particular, the decreasing rate of $\rho_h(t)$, affected by the constant $l_h$, introduces a lower bound on the speed of convergence of $e_h(t)$, $h \in \{ d, r_k, \alpha_k  \}, k=\{3,\ldots,n\}$. Furthermore, depending on the accuracy (resolution) of the sensors, the constants $\rho_{\infty,h}$ can be set arbitrarily small, thus achieving practical convergence of $e_h(t)$ to zero.  

The task is to synthesize  decentralized feedback control laws such that, given $	-\underline{b}_h \rho_{h}(0) < e_h(0) < \bar{b}_{h} \rho_{h}(0)$, $h \in \{ d, r_k, \alpha_k  \}, k=\{3,\ldots,n\}$, the formation errors $e_h(t)$ satisfy \eqref{eq:e_h_bound} for all $t\geq0$ leading to (practical) stabilization of the errors in \eqref{eq:eij}, \eqref{eq:r_error}, and \eqref{eq:alpha_error}.
%################################################################
\subsubsection{Selection of the Performance Bounds} \label{sec:bounds_select} We can incorporate the requirement of $\|{p}_{ji}(t)\|\neq 0, \forall(j,i) \in \mathcal{E}, \forall t\geq 0$ by choosing the maximum (absolute) values of the \textit{performance bounds} $-\underline{b}_h \rho_{h}(t), \bar{b}_{h} \rho_{h}(t)$ on $e_h(t)$, $h \in \{ d, r_k, \alpha_k  \}, k=\{3,\ldots,n\}$ in \eqref{eq:e_h_bound}, appropriately. In particular, from \eqref{eq:eij} and \eqref{eq:e_h_bound},
choosing $\underline{b}_d \rho_d(t)$ such that $\inf_{t\geq0} \left( (d_{21}^{\ast}(t))^2 - \underline{b}_d \rho_b(t)\right) \geq 0$ is sufficient to ensure $\|{p}_{21}(t)\| > 0$ for all $t\geq0$. On the other hand, $\bar{b}_d$ can be chosen arbitrarily without affecting the positivity of $\|{p}_{21}(t)\|$. Furthermore, as boundedness of $r_k$ in \eqref{eq:log_ratio} implies $\|{p}_{ki}(t)\|, \|{p}_{kj}(t)\| > 0$, from boundedness of $r_k^{\ast}$ in \eqref{eq:r_error} and the fact that $\max(\rho_{r_k}(t)) = \rho_{r_k}(0) = 1$, setting any bounded arbitrary values for $\underline{b}_{r_k}$ and $\bar{b}_{r_k}$ in \eqref{eq:e_h_bound} ensures $\|{p}_{ki}(t)\|, \|{p}_{kj}(t)\| > 0$, for $(k,i), (k,j) \in \mathcal{E} \setminus \{(2,1)\}$ and $\forall t \geq 0$. Moreover, notice that since the edge-angles are defined over the domain $\alpha_{kij} \in [0, 2\pi)$, from \eqref{eq:alpha_error}, by setting $\underline{b}_{\alpha_k} \leq \alpha_{kij}^{\ast}$ and $\bar{b}_{\alpha_k} \leq 2\pi - \alpha_{kij}^{\ast}$ in \eqref{eq:e_h_bound}, we can enforce this domain for the edge-angles, which avoids sudden changes from $2\pi$ to $0$. This further ensures continuous angle errors in \eqref{eq:alpha_error} leading to a smooth control action. Finally, notice that each agent can observe its initial formation errors and select $\underline{b}_h$ and $\bar{b}_{h}$, $h \in \{ d, r_k, \alpha_k  \}, k=\{3,\ldots,n\}$ in agreement with the above conditions to further ensure the requirement of $-\underline{b}_h \rho_{h}(0) < e_h(0) < \bar{b}_{h} \rho_{h}(0)$.

%################################################################
\subsubsection{Transformed Errors}
The problem of designing a controller that meets the error constraints in \eqref{eq:e_h_bound} can be transformed into establishing the boundedness of certain modulated error signals \cite{bechlioulis2008robust,bechlioulis2014low}. More specifically, to handle the time-varying constraints in \eqref{eq:e_h_bound}, a time-varying error transformation technique will be used to convert each of the original error dynamics $\dot{e}_d$, $\dot{e}_{r_{k}}$, and $\dot{e}_{\alpha_{k}}$ (given in \eqref{eq: dot_e21}, \eqref{eq:err_dyn}, respectively) under the constraints \eqref{eq:e_h_bound} into equivalent unconstrained ones, whose stability merely ensures satisfaction of the constraints given in \eqref{eq:e_h_bound}. First, we define the modulated formation errors as:
\begin{equation} \label{modu_err}
	\tilde{e}_h(t) \coloneqq \dfrac{e_h(t)}{\rho_h(t)}, \quad h \in \{ d, r_k, \alpha_k  \}, k=\{3,\ldots,n\}.
\end{equation}
To transform the constrained error dynamics (in the sense of \eqref{eq:e_h_bound}) into an equivalent unconstrained one, we introduce the following error transformation:
\begin{equation}\label{eq:mappings}
		\sigma_h = T_h(\tilde{e}_h), \quad h \in \{ d, r_k, \alpha_k  \}, k=\{3,\ldots,n\},
\end{equation}
where $\sigma_h, h \in \{ d, r_k, \alpha_k  \}, k=\{3,\ldots,n\}$, are the transformed errors corresponding to $e_h$. Moreover, $T_h(.) : (- \underline{b}_h , \bar{b}_h) \rightarrow (- \infty , + \infty)$, denote smooth, \textit{strictly increasing} \textit{bijective} mappings satisfying $T_h(0)= 0$. Note that $e_h = 0$ if and only if $\sigma_h = 0$. Finally, notice that maintaining the boundedness of $\sigma_h(t)$, enforces $-\underline{b}_h<\tilde{e}_h(t)<\bar{b}_h$, and consequently the satisfaction of \eqref{eq:e_h_bound}. Taking the time derivatives of \eqref{eq:mappings}, yields:
\begin{equation}\label{eq: derivat_map_k}
	\dot{\sigma}_h =  \xi_h (\dot{e}_h - \tilde{e}_h \, \dot{\rho}_h ), \;\; h \in \{ d, r_k, \alpha_k  \}, k=\{3,\ldots,n\},
\end{equation}
where
\begin{equation}\label{derivative_coef}
	\begin{gathered}
		\xi_h \coloneqq \frac{1}{\rho_h(t)} \frac{\partial T_h(\tilde{e}_h)}{\partial \tilde{e}_h} > 0, \\
		h \in \{ d, r_k, \alpha_k  \}, k=\{3,\ldots,n\}. 
	\end{gathered}
\end{equation}

In the sequel, we shall consider the following logarithmic function as a proper choice for the mapping functions in \eqref{eq:mappings}: 
	\begin{equation}	\label{eq:proper_map}
			\sigma_h = T_h(\tilde{e}_h) = \ln \left( \dfrac{\bar{b}_h \tilde{e}_h + \bar{b}_h \underline{b}_h}{\bar{b}_h \underline{b}_h - \underline{b}_h \tilde{e}_h} \right), 
	\end{equation}
	where $h \in \{ d, r_k, \alpha_k  \}, k=\{3,\ldots,n\}$. Note that, the specific form in \eqref{eq:proper_map} satisfies the aforementioned properties for $T_h(.)$ and $\tilde{e}_h \in  (- \underline{b}_h , \bar{b}_h)$ if and only if $\sigma_h \in (- \infty , + \infty)$. 
	
\begin{remark}[PPC Design Philosophy]
	When $-\underline{b}_h \rho_h(0) < e_h(0) < \bar{b}_h \rho_h(0)$, $h \in \{ d, r_k, \alpha_k  \}, k=\{3,\ldots,n\}$, based on the properties of the error transformations \eqref{eq:mappings}, prescribed performance in the sense of \eqref{eq:e_h_bound} is achieved, if $\sigma_h(t)$ are kept bounded. Notice that, although for $\sigma_h \in \mathbb{R}$ the prescribed performance bounds in \eqref{eq:e_h_bound} are satisfied, the boundedness of $\sigma_h$ is required to guarantee well-defined bounded control inputs. Moreover, it is important to note that the specific bounds of $\sigma_h$ (no matter how large they are, which is the key property of the adopted error transformation) do not affect the achieved transient and steady-state performance on $e_h(t)$, which is solely determined by \eqref{eq:e_h_bound} and thus by the selection of the performance functions $\rho_h(t)$ as well as the scaling constants $\bar{b}_h$ and $\underline{b}_h$.
\end{remark}

%################################################################
\subsubsection{Proposed Control Laws} \label{subss}
The following lemma is useful for the control design and stability analysis.
\begin{lemma} \label{lem:bipolar_basis_global}
	For a given triangular directed sub-graph as in Fig.\ref{fig:trig_framework}, the bipolar coordinates basis $\widehat{r}_k$, $\widehat{\alpha}_k$ (see Fig.\ref{fig:local_bipolar}) associated with the virtual Cartesian frame $\left\lbrace C_k \right\rbrace$ in Fig.\ \ref{fig:local_cart} can be expressed with respect to the global coordinate system as follows:
	\begin{subequations} \label{eq:bipolar_basis_in_globalframe}
		\begin{align}
			\widehat{\alpha}_{k} &= - f_1(r_k, \alpha_{kij}) z_{ji} + f_2(r_k, \alpha_{kij}) J^T z_{ji}\;, \label{eq:alpha_hat} \\
			\widehat{r}_k  &= f_2(r_k, \alpha_{kij}) z_{ji} + f_1(r_k, \alpha_{kij}) J^T z_{ji}\;, \label{eq:r_hat}
		\end{align}
	\end{subequations}
	for $k\geq 3$, $(k,i), (k,j) \in \mathcal{E} \setminus \{(2,1)\}$, $i<j<k$, where $z_{ji}$ is the bearing vector associated with edge $(j,i)$, $J^T$ is the $\frac{\pi}{2}$ clockwise rotation matrix, and
	\begin{subequations}\label{eq:B1_B2}
		\begin{align}
			f_1(r_k, \alpha_{kij}) &= \dfrac{- \sinh r_k \sin \alpha_{kij}}{\cosh r_k - \cos \alpha_{kij}}, \\
			f_2(r_k, \alpha_{kij}) &= \dfrac{\cos \alpha_{kij} \cosh r_k - 1}{\cosh r_k - \cos \alpha_{kij}}.
		\end{align}
	\end{subequations}
\end{lemma}
\begin{IEEEproof}
	See Appendix \ref{appen:proof_lem_bipolar_basis}.
\end{IEEEproof}

Notice that in the proposed formation control setup the leader (agent 1) does not participate in forming the desired shape, thus its behaviour is independent from the other agents. In this respect, the leader's control law $u_L(t)$ is designed  for objectives such as trajectory tracking, position stabilization, etc., in the presence of external disturbances/uncertainties $\delta_1(t)$. Note that, the response of the leader should be stable, therefore, in the following we will assume $u_L(t)$ is uniformly bounded and designed such as to achieve a desired high-level formation coordination task.

We propose the following formation control laws:
\begin{subequations} \label{eq:control_law}
	\begin{align}
		u_1 &= u_L(t) \label{eq:cont1}\\
		u_2 &= \xi_d \, \sigma_d \, {p}_{21} \label{eq:cont2} \\
		u_k &= - \xi_{r_k} \, \sigma_{r_k} \, \widehat{r}_{k} -\xi_{\alpha_k} \, \sigma_{\alpha_k} \, \widehat{\alpha}_{k} , \quad k=3,\dots,n, \label{eq:cont_k}
	\end{align}
\end{subequations}
where $\widehat{\alpha}_{k}$ and $\widehat{r}_{k}$ are the bipolar coordinates basis associated with agent $k\geq 3$ obtained in \eqref{eq:bipolar_basis_in_globalframe}, and from \eqref{derivative_coef} and \eqref{eq:mappings}, $\xi_h$,  $h \in \{ d, r_k, \alpha_k  \}, k=\{3,\ldots,n\}$ is given by:
\begin{equation}\label{eq:psi,xi}
	\xi_h = \dfrac{1}{\rho_h(t)} \left(\dfrac{1}{\tilde{e}_h + \underline{b}_h} - \dfrac{1}{\tilde{e}_h - \bar{b}_h}\right), 
\end{equation}
which is lower bounded by a positive constant over its domain $\tilde{e}_h  \in (-\underline{b}_h, \bar{b}_{h})$ owing to strict positiveness of $\rho_h(t)$.
\begin{remark}
	The control law \eqref{eq:cont_k} indicates that the motion of the follower $k\geq3$ results by the superposition of the motions along each of the orthogonal bipolar coordinate basis $\widehat{r}_{k}, \widehat{\alpha}_{k} \in \mathbb{R}^2$ to compensate the formation errors $e_{r_{k}}, e_{\alpha_{k}}$ with the given constraints in \eqref{eq:e_h_bound}. In this respect, notice that one can select the performance bounds (error constraints) on $e_{r_{k}}, e_{\alpha_{k}}$ in \eqref{eq:e_h_bound} arbitrarily without any constraint infeasibility issues since these two error variables can vary independently along their respective bipolar coordinate basis (i.e., $e_{r_{k}}, e_{\alpha_{k}}$ are not interdependent). Moreover, notice that for implementing \eqref{eq:cont_k}, from \eqref{eq:bipolar_basis_in_globalframe}, agent $k$ should know $z_{ji}$, which is the relative bearing between its neighbours $i<j \in \mathbb{N}$. We argue that agent $k$ can obtain $z_{ji}$ by direct measurements of $z_{ki}, z_{kj}$, and the ratio of the distances $r_{kij}, i<j<k$, which are available. First, notice that $p_{ji} = p_{ki} - p_{kj} = \|p_{ki}\| z_{ki} - \|p_{kj}\| z_{kj}$. Let $z_k \coloneqq r_{kij} z_{ki} - z_{kj} \in \mathbb{R}^2$. One can verify that $z_k$ is parallel with $p_{ji}$; consequently, normalizing $z_k$ gives $z_{ji}$.
\end{remark} 

\begin{remark} \label{re:coordin_free} 
	Although the proposed control laws \eqref{eq:control_law} are given with respect to a global coordinate frame (only for the sake of analysis), we emphasize that the proposed formation controller can be implemented in any arbitrarily oriented local coordinate frame (i.e., in a coordinate-free fashion). First, notice that according to the leader's objective (e.g., going to a specific position or following a trajectory, etc.), since it is not involved in the process of generating the desired shape, the leader can perform the required calculations for its control law with respect to its own local coordinate frame. Second, let $g_h(e_h) \coloneqq \xi_h \sigma_h, h \in \{ d, r_k, \alpha_k  \}, k=\{3,\ldots,n\}$ in \eqref{eq:cont2}, \eqref{eq:cont_k}, where all $g_h(e_h)$ are scalar functions of the formation errors. Now let the superscript $[k], k\geq 2$, indicates a quantity expressed in the local coordinate frame of the $k$-th agent. Furthermore, suppose that $\mathcal{R}_k \in \mathrm{SO}(2)$ is the transformation (rotation) matrix from the $k$-th local frame to the global frame. Notice that, we have $u_k=\mathcal{R}_k u_k^{[k]}$, $p_{ki}=\mathcal{R}_k p_{ki}^{[k]}=\mathcal{R}_k(p_i^{[k]}-p_k^{[k]})$, and consequently from \eqref{eq:bearing} we get $z_{ki}=\mathcal{R}_k z_{ki}^{[k]}, i<k \in \mathbb{N}$. Considering \eqref{eq:cont_k}, we have:
	\begin{align}
		u_k^{[k]}=\mathcal{R}_k^{-1}u_k &= \mathcal{R}_k^{-1} \left( -g_{r_{k}}(e_{r_k}) \widehat{r}_{k} - g_{\alpha_{k}}(e_{\alpha_k}) \widehat{\alpha}_{k} \right) \nonumber \\
		&= -g_{r_{k}}(e_{r_k}) \mathcal{R}_k^{-1} 	\widehat{r}_{k} - g_{\alpha_{k}}(e_{\alpha_k}) \mathcal{R}_k^{-1} \widehat{\alpha}_{k} \nonumber \\
		&= -g_{r_{k}}(e_{r_k}) \widehat{r}_{k}^{\,[k]} - g_{\alpha_{k}}(e_{\alpha_k}) \widehat{\alpha}_{k}^{\,[k]}, \label{eq:u_k_local coordi}
	\end{align}
	where from \eqref{eq:bipolar_basis_in_globalframe} we get:
	\begin{subequations}\label{eq:bipolar_basis_localfram}
		\begin{align}
			\widehat{\alpha}_{k}^{\,[k]} = \mathcal{R}_k^{-1} \widehat{\alpha}_{k} &= -f_1 \mathcal{R}_k^{-1} z_{ji} + f_2 J^T \mathcal{R}_k^{-1} z_{ji} \nonumber \\
			&= -f_1 z_{ji}^{[k]} + f_2 J^T z_{ji}^{[k]},  \\
			\widehat{r}_{k}^{\,[k]} = \mathcal{R}_k^{-1} \widehat{r}_{k} &= f_2 \mathcal{R}_k^{-1} z_{ji} + f_1 J^T \mathcal{R}_k^{-1} z_{ji} \nonumber \\
			&= f_2 z_{ji}^{[k]} + f_1 J^T z_{ji}^{[k]},
		\end{align}
	\end{subequations}
	in which the permutation property between $\mathcal{R}_k^{-1}$ and $J^T$ is employed to achieve the right-hand sides (since both of them are rotation matrices). Note that, the values of the scalar functions $f_1(\alpha_{kij}, r_k), f_2(\alpha_{kij}, r_k)$ as well as $g_h(e_h) \coloneqq \xi_h \sigma_h, h \in \{ d, r_k, \alpha_k  \}, k=\{3,\ldots,n\}$ do not depend on the coordinate systems since their arguments (i.e., edge-angles, logarithm of ratio of distances, and their errors) are the same in any coordinate system. One can verify  that \eqref{eq:bipolar_basis_localfram} and \eqref{eq:u_k_local coordi} have the same form as \eqref{eq:bipolar_basis_in_globalframe} and \eqref{eq:cont_k}, respectively, where all the quantities are expressed
	with respect to the $k$th local coordinate frame. This indicates that the decentralized control law \eqref{eq:cont_k} can be implemented in arbitrarily oriented local coordinate frame of agent $k$. The same claim can be verified in  a similar manner for the control law of agent 2 (secondary leader) in \eqref{eq:cont2}.
\end{remark}

%################################################################
\subsection{Stability Analysis}
The main results of this work are summarized in the
following theorems. Theorem \ref{th:agent2} indicates that the secondary leader (i.e., agent 2) can keep a certain (in general time-varying) distance with respect to the leader (which can move freely). Compensation of the formation errors assigned to the followers (agents $k\geq3$) are derived in Theorem \ref{th:agent_k}.

\begin{theorem} \label{th:agent2}
	Consider agents 1 and 2 with dynamics \eqref{eq:singledyn} and a desired formation given by a directed graph $\mathcal{G} = (\mathcal{V},\mathcal{E})$ under Assumption \ref{assu:G} with a desired (time-varying) distance $\inf_{t\geq0} \left(d_{21}^{\ast}(t)\right)>0$ between agents 1 and 2. Given that $-\underline{b}_{d} \rho_{d}(0) < e_d(0) < \bar{b}_{d} \rho_{d}(0)$, where $\underline{b}_d \rho_d(t)$ and $\bar{b}_d$ are chosen as explained in Section \ref{sec:bounds_select}, the decentralized control protocols \eqref{eq:cont1}, \eqref{eq:cont2} guarantee $-\underline{b}_{d} \rho_{d}(t) < e_d(t) < \bar{b}_{d} \rho_{d}(t)$ and $\|p_{21}(t)\| > 0$, for all $t\geq0$ as well as boundedness of all closed-loop signals.
\end{theorem}
\begin{IEEEproof}
	 The proof is provided in Appendix \ref{appen:th1}. The proof proceeds in three phases: First, we show that $\tilde{e}_{d}(t)$ remains within $(-\underline{b}_{d}, \bar{b}_{d})$ for a specific time interval $[0, \tau_{\mathrm{2,max}})$ (i.e., the existence and uniqueness of a maximal solution) by noting that the closed-loop system of $\dot{\tilde{e}}_d$ is Lipschitz continuous. Next, by establishing boundedness of the mapped error $\sigma_d$ we prove that the proposed control scheme guarantees, for all $[0, \tau_{\mathrm{2,max}})$: a) the boundedness of all closed loop signals as well as b) that $\tilde{e}_{d}(t)$ remains strictly in a compact subset of $(-\underline{b}_{d}, \bar{b}_{d})$, which leads to $\tau_{\mathrm{2,max}}=\infty$ (i.e., forward completeness), thus finalizing the proof.
\end{IEEEproof}

\begin{theorem} \label{th:agent_k}
		Consider a group of $n$ agents with dynamics \eqref{eq:singledyn} in a 2-D plane. Let the desired formation be given by a directed graph $\mathcal{G} = (\mathcal{V},\mathcal{E})$ under Assumption \ref{assu:G} along with the sets of  logarithms of the desired ratio of distances $r_k^{\ast}$ and desired edge-angles $\alpha_{kij}^{\ast}$, $(k,i), (k,j) \in \mathcal{E} \setminus \{(2,1)\}$, $i<j<k$, assigned to agents $3\leq k\leq n$. Assume that $-\underline{b}_{h} \rho_{h}(0) < e_h(0) < \bar{b}_{h} \rho_{h}(0)$, $h \in \{ r_k, \alpha_k  \}, k=\{3,\ldots,n\}$, where $\underline{b}_h$ and $\bar{b}_h$ are chosen as in Section \ref{sec:bounds_select}. Under the stability results of Theorem \ref{th:agent2}, the decentralized control protocol \eqref{eq:cont_k} guarantees,  $-\underline{b}_{h} \rho_{h}(t) < e_h(t) < \bar{b}_{h} \rho_{h}(t)$ and $\|p_{ki}(t)\|, \|p_{kj}(t)\| > 0$, $(k,i), (k,j) \in \mathcal{E} \setminus \{(2,1)\}$, for all $t\geq0$, as well as boundedness of all closed-loop signals. 
\end{theorem}
\begin{IEEEproof}
	  See Appendix \ref{appen:th2} for the proof. The proof follows an organization similar to that of Theorem \ref{th:agent2}. In particular, first, we establish the results for agent 3's formation errors (i.e., for the first triangular subgraph of the desired formation). Next, by leveraging the established results for agent 3 and Theorem 1, as well as exploiting the hierarchical leader-follower structure of the formation control system, we extend the results to all of the agents by induction.
\end{IEEEproof}

\begin{remark}
	We highlight that our results in Theorems \ref{th:agent2} and \ref{th:agent_k} indicate \textit{almost} global convergence to the desired formation. In particular, satisfaction of  $-\underline{b}_{d} \rho_{d}(0) < e_d(0) < \bar{b}_{d} \rho_{d}(0)$ in Theorem \ref{th:agent2} as well as  satisfaction of $-\underline{b}_{r_k} \rho_{r_k}(0) < e_{r_k}(0) < \bar{b}_{r_k} \rho_{r_k}(0)$ in Theorem \ref{th:agent_k} along with choosing $\underline{b}_d \rho_d(t)$, $\bar{b}_d$, $\underline{b}_{r_k}$, and $\bar{b}_{r_k}$ according to Section \ref{sec:bounds_select} require agents not to be collocated initially with their neighbors (i.e., $\|p_{21}(0)\| \neq 0, \|p_{ki}(0)\| \neq 0, \|p_{kj}(0)\| \neq 0, (k,i), (k,j) \in \mathcal{E} \setminus \{(2,1)\} , i<j<k$). Moreover, satisfaction of $-\underline{b}_{\alpha_k} \rho_{\alpha_k}(0) < e_{\alpha_k}(0) < \bar{b}_{\alpha_k} \rho_{\alpha_k}(0)$ in Theorem \ref{th:agent_k} along with choosing $\underline{b}_{\alpha_k}$ and $\bar{b}_{\alpha_k}$ according to Section \ref{sec:bounds_select} requires $0 < \alpha_{kij}(0) < 2\pi$ which affects the acceptable initial positions for agent $k\geq3$ with respect to its neighbors, i.e., agent $k$ should not be initially collinear with agents $i$ and $j$ while locating on left- or right-hand side of them (being in the middle is feasible). Additionally, note that the target formation should respect the aforementioned conditions as well. Finally, we argue that the above restrictions only constitute a zero-measure set to avoid global convergence.
\end{remark}
%################################################################
\subsection{Formation Control with Orientation Adjustment} \label{subsec:orientation}
Following \eqref{eq:lemma_bipol}, owing to the above results and the proposed control laws in \eqref{eq:control_law}, the leader determines the position of the formation (e.g., by tracking a reference velocity/trajectory that leads to formation maneuvering), the secondary leader determines the formation scale by tracking a time-varying desired distance $d_{21}^{\ast}(t)$, and the followers contribute to obtain the desired shape based on the defined errors in bipolar coordinates. In this section, we shall extend the aforementioned results to formation control with orientation adjustment, where, by using an extended control law for the secondary leader, we can obtain a certain (in general, time-varying) desired orientation for the formation as well. 

\begin{figure}[tb]
	\flushleft
	\includegraphics[scale=0.43]{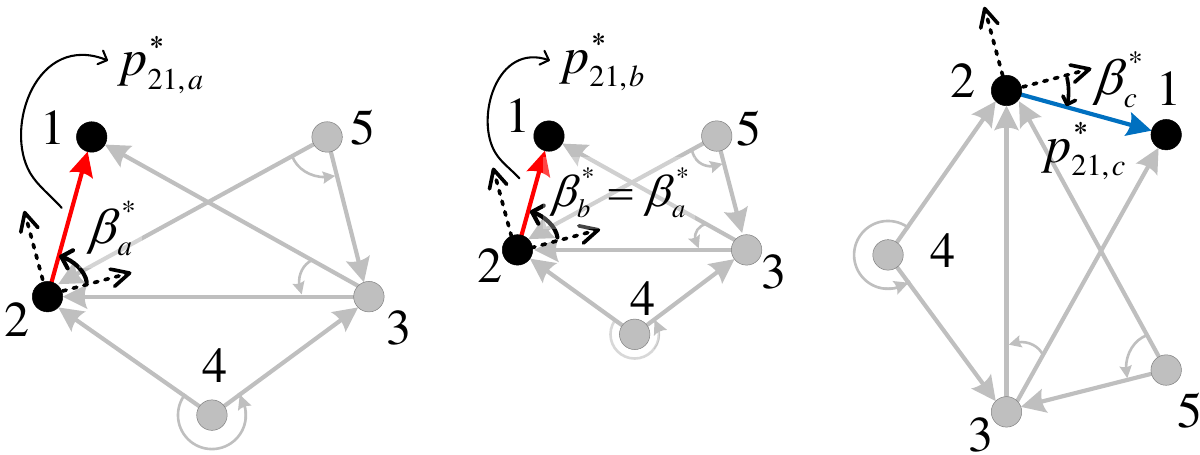}
	\caption{ Given a desired sensing graph $\mathcal{G}$ as in Fig.\ref{fig:angle_example}, in each case the desired formation is characterized by different desired relative positions between agents 2 and 1, whereas the sets of desired edge-angles and ratio of the distances for followers ($i\geq3$) are the same. The dashed arrows show the local coordinate frame of agent 2 in which the formation orientation can be characterized by the desired bearing angle $\beta^{\ast}$. $p_{21,a}^{\ast}$ and $p_{21,b}^{\ast}$ have the same orientation but different length while $p_{21,a}^{\ast}$ and $p_{21,c}^{\ast}$ have different orientations with the same length.}
	\label{fig:orient}
\end{figure}

Let  $\beta \coloneqq \mathrm{arctan2}(z_{21,y}, z_{21,x}) \in (- \pi, \pi]$ be the bearing angle between agents 2 and 1, where $z_{21} = [z_{21,x}, z_{21,y}]^T$ denotes the corresponding bearing vector\footnote{$\mathrm{arctan2}$ is the two argument arc tangent function \cite[Appendix A]{spong2020robot}.}. Now consider the following new control requirement in addition to the ones in \eqref{eq:lemma_bipol}:
\begin{equation} \label{eq:desired_bearing_angle}
	\beta(t) \rightarrow {\beta}^{\ast}, \quad \mathrm{as} \quad t \rightarrow \infty
\end{equation}
where $\beta^{\ast}$ is a desired bearing angle associated with a desired bearing vector $z_{21}^{\ast}$ between agents 2 and 1. Note that $\beta^{\ast}$ determines the desired orientation for the formation. In this regard, $\beta^{\ast}$ and $d_{21}^{\ast}$ determine a desired relative position vector $p_{21}^{\ast}$ between agents 2 and 1, since $p_{21}^{\ast} = d_{21}^{\ast} z_{21}^{\ast}$. Indeed satisfaction of \eqref{eq:2nd_agent_dist} and \eqref{eq:desired_bearing_angle} is equivalent to: $p_2(t)  - p_1(t) \rightarrow {p}_{21}^{\ast}$, as $t \rightarrow \infty$, see Fig.\ref{fig:orient} for further illustrations. In this way the secondary leader not only controls the scaling of the desired formation at the steady-state, but also alters the formation orientation by modifying its bearing angle with respect to the leader. In other words, satisfaction of \eqref{eq:desired_bearing_angle} along with \eqref{eq:lemma_bipol} achieves the desired formation only up to translations. Note that, given a ${p}_{21}^{\ast}$ in a global coordinate system, one can always find $d_{21}^{\ast}$ and $\beta^{\ast}$. Having ${p}_{21}^{\ast}$ defined in a global coordinate system is not restrictive since it is only required to be accessible to agent 2, hence, as it is shown in Fig.\ref{fig:orient}, one can always consider agent 2's local coordinate frame as the reference frame in which the formation orientation is determined.
Let us define the bearing angle error between agent 2 and 1 as:
\begin{equation} \label{eq:e_beta}
	e_{\beta} = \beta(t) - \beta^{\ast}(t),
\end{equation}
where $\beta^{\ast}(t): \mathbb{R} \rightarrow (- \pi, \pi)$ is a continuously differentiable function of time with a bounded derivative representing the desired (in general, time-varying) orientation between agents 2 and 1 (which is equivalent to the desired formation orientation). Similarly to Section \ref{Sec:Ctrl_Design_Analysis}, the PPC method can be adopted to design a robust control law for practical stabilization of $e_{\beta}$ such that:
\begin{equation} 
	-\underline{b}_{\beta} \rho_{\beta}(t)  < e_{\beta}(t) < \bar{b}_{\beta} \rho_{\beta}(t) \label{eq:beta_const}
\end{equation}
by utilizing an unconstrained transformed error $\sigma_{\beta} = T_{\beta}(\tilde{e}_{\beta})$, where $\underline{b}_{\beta}, \bar{b}_{\beta}>0$, and $\rho_{\beta}$, $\tilde{e}_{\beta}$, $\sigma_{\beta}$, $\xi_{\beta}$ are defined similarly to \eqref{pf_rho}, \eqref{modu_err}, \eqref{eq:proper_map}, and \eqref{eq:psi,xi}, respectively. Moreover, akin to Section \ref{sec:bounds_select}, to ensure $\beta \in (- \pi, \pi)$ and avoid potential singularities (i.e., $\beta = \pi$ or $-\pi$) we should have $\underline{b}_{\beta} \rho_{\beta}(t) \geq \pi + \beta^{\ast}(t) $ and $\bar{b}_{\beta} \rho_{\beta}(t) \leq \pi - \beta^{\ast}(t)$ for all $t \geq 0$. Hence, $\inf_{t\geq0} \left(\pi + \beta^{\ast}(t) - \underline{b}_{\beta} \rho_{\beta}(t)\right) \leq 0$ and $\inf_{t\geq0} \left(\pi - \beta^{\ast}(t) - \bar{b}_{\beta} \rho_{\beta}(t)\right) \geq 0$ are sufficient to ensure $\beta \in (- \pi, \pi)$ for all $t\geq0$. We thus propose the following extended control law for agent 2:
\begin{equation}
	u_2 = \xi_d \, \sigma_d \, {p}_{21} + \xi_{\beta} \sigma_{\beta} J z_{21} \label{eq:cont2_extend}
\end{equation}
which indicates that the motion of secondary leader results from the superposition of the motions along orthogonal directions $z_{21}$ and $Jz_{21} = z_{21}^{\bot}$ to compensate the distance error $e_d$ and the bearing angle error $e_{\beta}$, with the given constraints in \eqref{eq:e_h_bound} and \eqref{eq:beta_const}, respectively. 

The aforementioned results are summarized in Corollary \ref{coro:1}, which along with Theorem \ref{th:agent_k} extends the results to formation control with scaling and orientation adjustment.

\begin{corollary} \label{coro:1}
	Consider agents 1 and 2 with dynamics \eqref{eq:singledyn} and a desired formation given by a directed graph $\mathcal{G} = (\mathcal{V},\mathcal{E})$ under Assumption \ref{assu:G} with a desired (time-varying) distance $\inf_{t\geq0} \left(d_{21}^{\ast}(t)\right)>0$ and a desired bearing angle $-\pi < \beta^{\ast}(t) < \pi$ between agents 1 and 2. Given $-\underline{b}_{h} \rho_{h}(0) < e_h(0) < \bar{b}_{h} \rho_{h}(0)$, $h \in \{ d, \beta\}$, where $\underline{b}_h \rho_{h}(t)$ and $\bar{b}_h$ are chosen as explained in Sections \ref{sec:bounds_select} and \ref{subsec:orientation}, the decentralized control protocols \eqref{eq:cont1} and \eqref{eq:cont2_extend} guarantee $-\underline{b}_{h} \rho_{h}(t) < e_h(t) < \bar{b}_{h} \rho_{h}(t)$ and $\|p_{21}(t)\| > 0$, for all $t\geq0$ as well as boundedness of all closed-loop signals.
\end{corollary}
\begin{IEEEproof}
 The proof is similar to Theorem \ref{th:agent2} and thus is omitted for brevity. Notice that based on the control law \eqref{eq:cont2_extend}, $e_d$ and $e_{\beta}$ can also be independently treated (due to orthogonality of the control directions).
\end{IEEEproof}

%################################################################
\section{Simulations Results} 
\label{Sec:Simu}  

\begin{figure*}[htb]
	\centering
	\includegraphics[width=\textwidth]{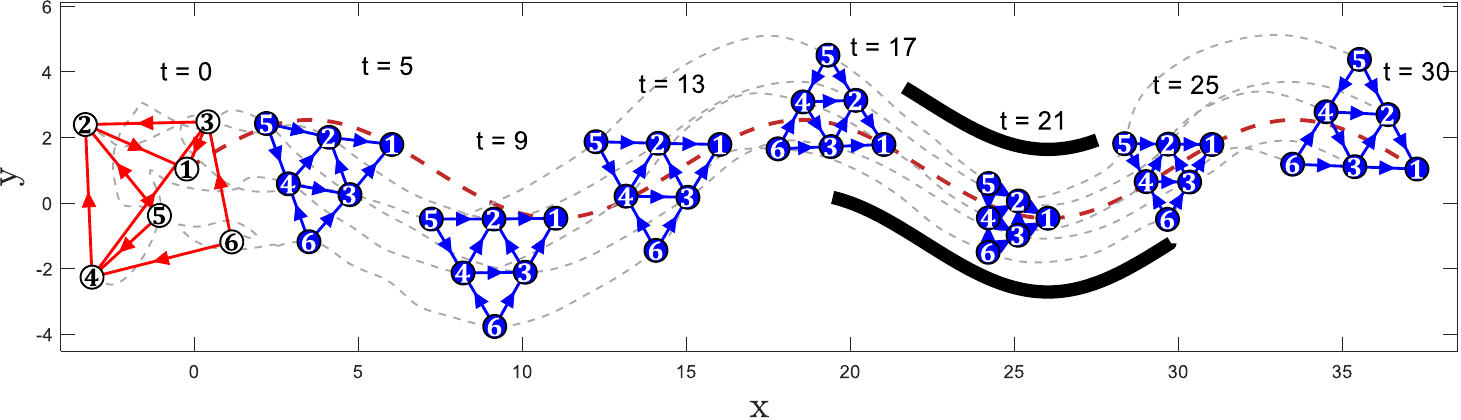}
	\caption{Starting from arbitrary initial positions, agents converge to the desired shape while following the leader's (agent 1's) motion. The scale and orientation of the formation is adjusted by agent 2 along the way. In particular, roughly around $t = 14$ agent 2 starts following a time-varying desired bearing and distance w.r.t. agent 1 that leads the formation to pass through a narrow passage (black curves).}
	\label{fig:Maneu}
\end{figure*}

\begin{figure}[htb]
	\centering
	%	\flushleft
	\begin{subfigure}[t]{0.225\textwidth}
		\centering
		\includegraphics[width=\textwidth]{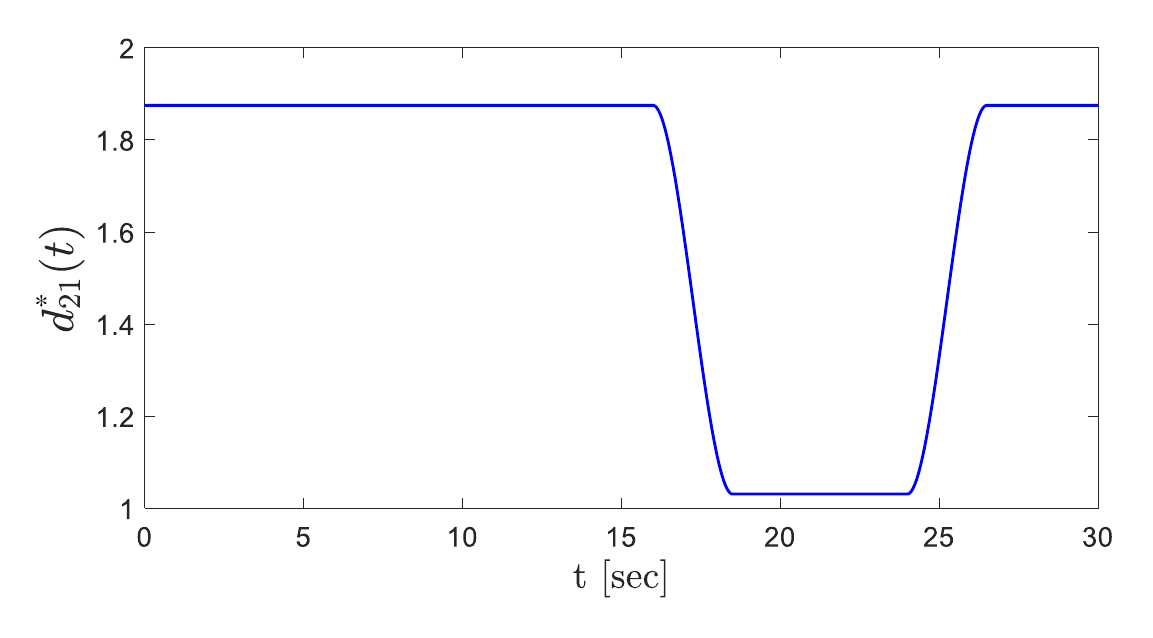}
	\end{subfigure}
	~
	\begin{subfigure}[t]{0.225\textwidth}
		\centering
		\includegraphics[width=\textwidth]{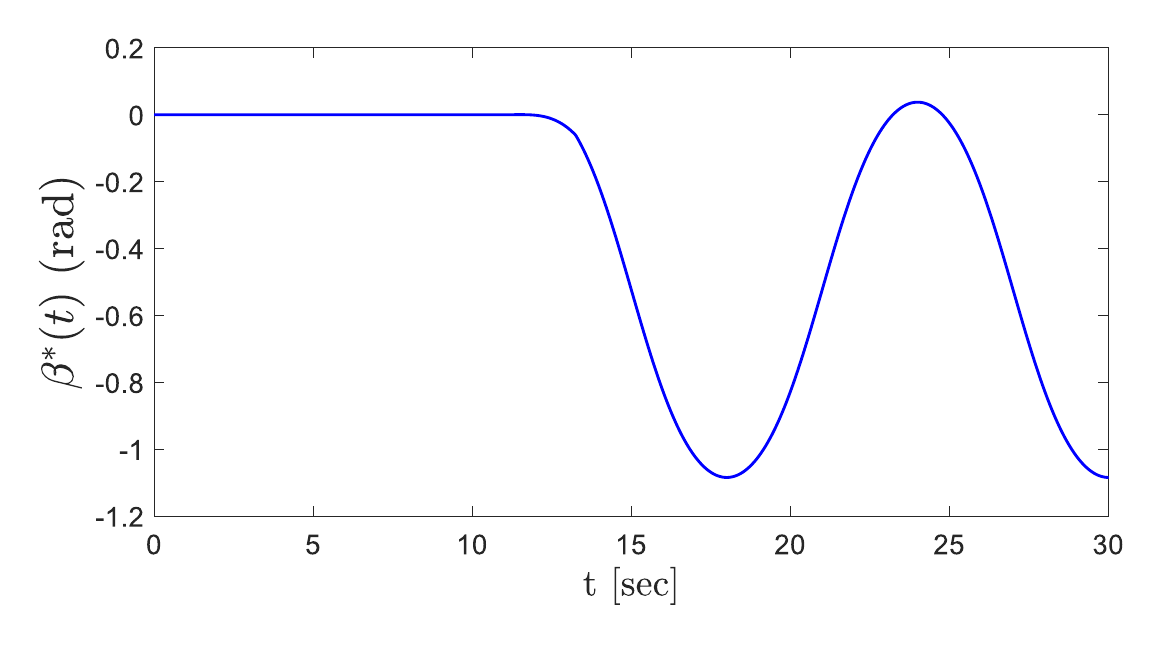}
	\end{subfigure}
	\caption{Agent 2's desired (time-varying) distance $d_{21}^{\ast}(t)$ and bearing angle $\beta^{\ast}(t)$.}
	\label{fig:dist_beta_refs}
\end{figure}

\begin{figure}[ht]
	\centering
	%	\flushleft
	\begin{subfigure}[t]{0.235\textwidth}
		\centering
		\includegraphics[width=\textwidth]{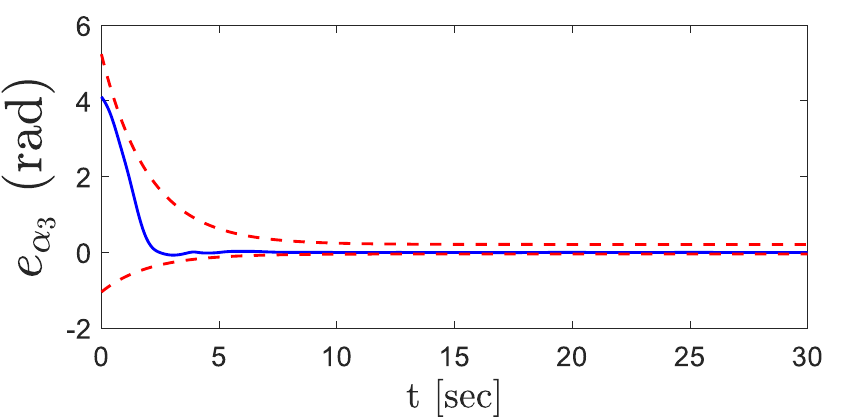}
	\end{subfigure}
	~
	\begin{subfigure}[t]{0.235\textwidth}
		\centering
		\includegraphics[width=\textwidth]{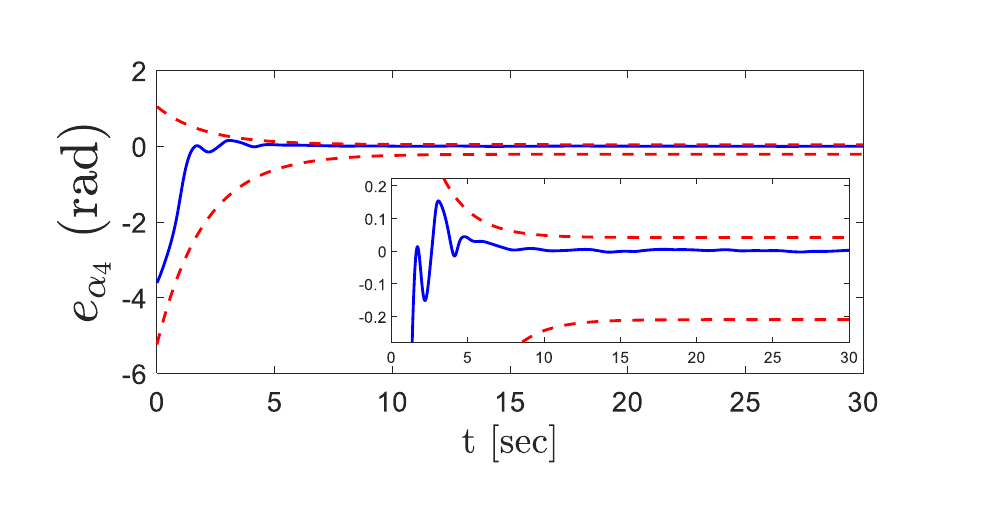}
	\end{subfigure}
	~
	\begin{subfigure}[t]{0.235\textwidth}
		\centering
		\includegraphics[width=\textwidth]{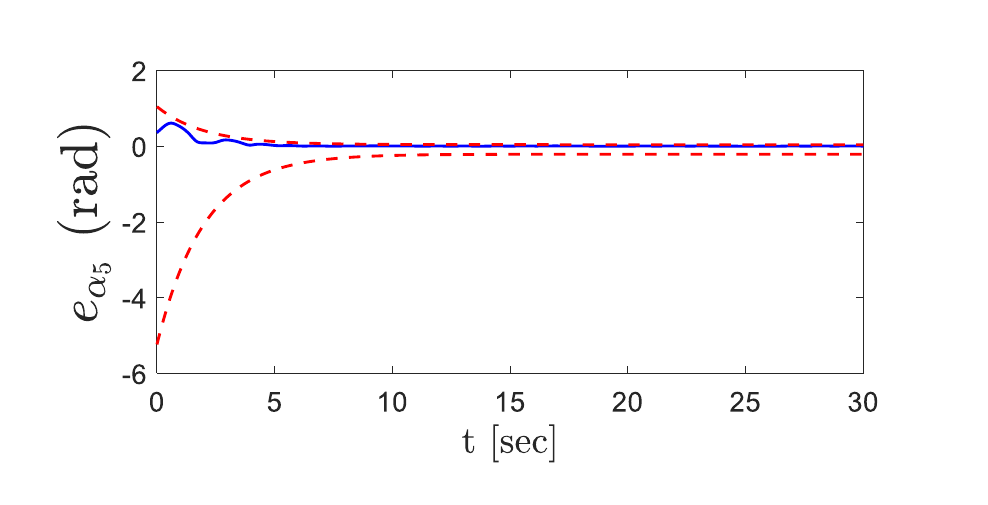}
	\end{subfigure}
	~
	\begin{subfigure}[t]{0.235\textwidth}
		\centering
		\includegraphics[width=\textwidth]{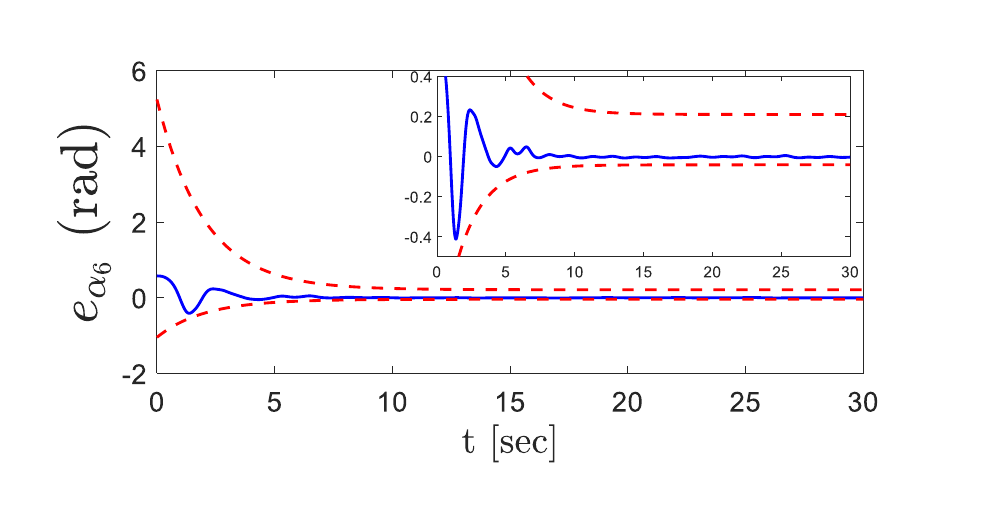}
	\end{subfigure}
	\caption{Evolution of the edge-angle errors $\alpha_k, k\geq 3$. The magnified subplots provide the details of error evolution in the steady-state.}
	\label{fig:e_alpha}
\end{figure}

\begin{figure}[ht]
	\centering
	%	\flushleft
	\begin{subfigure}[t]{0.235\textwidth}
		\centering
		\includegraphics[width=\textwidth]{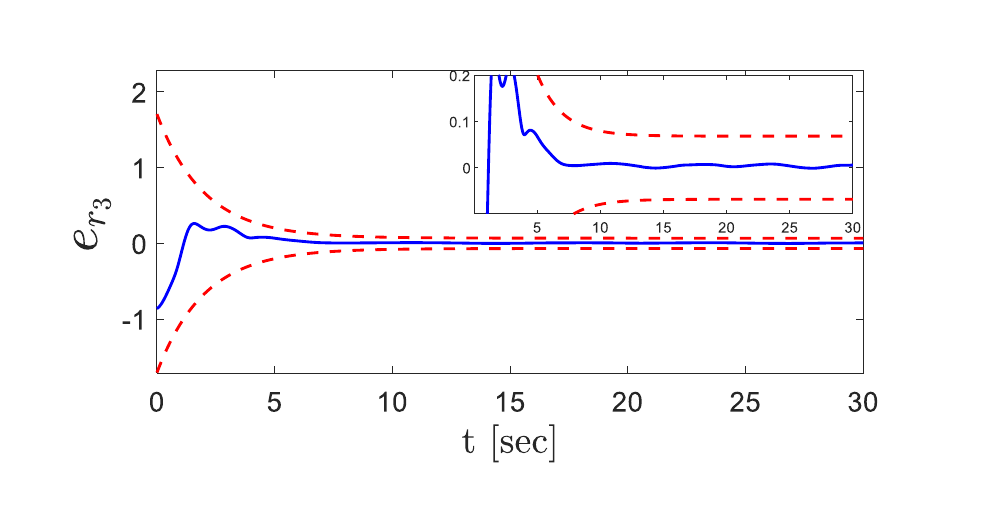}
	\end{subfigure}
	~
	\begin{subfigure}[t]{0.235\textwidth}
		\centering
		\includegraphics[width=\textwidth]{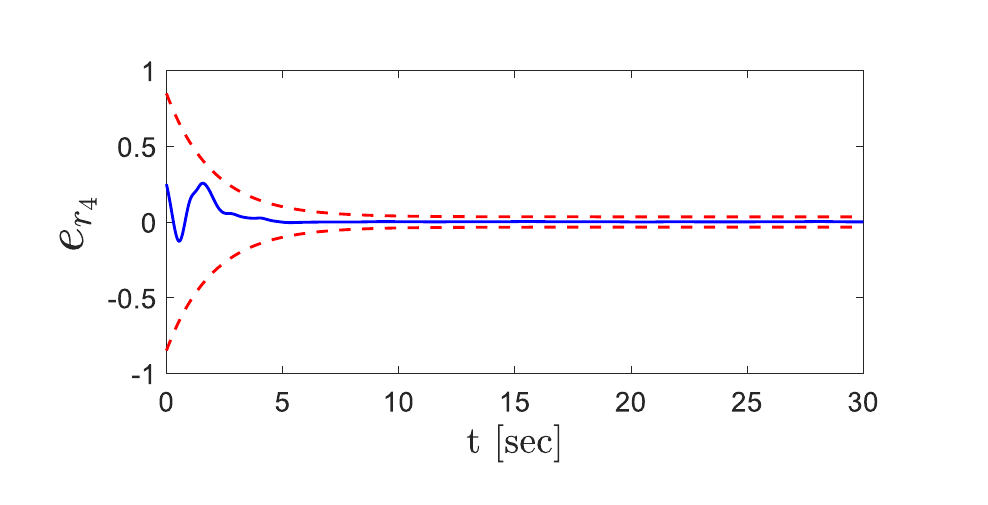}
	\end{subfigure}
	~
	\begin{subfigure}[t]{0.235\textwidth}
		\centering
		\includegraphics[width=\textwidth]{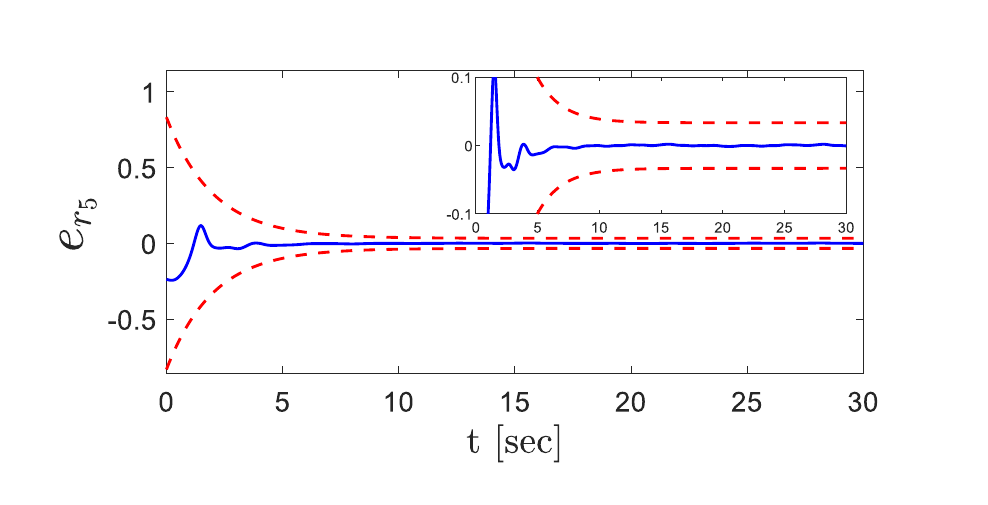}
	\end{subfigure}
	~
	\begin{subfigure}[t]{0.235\textwidth}
		\centering
		\includegraphics[width=\textwidth]{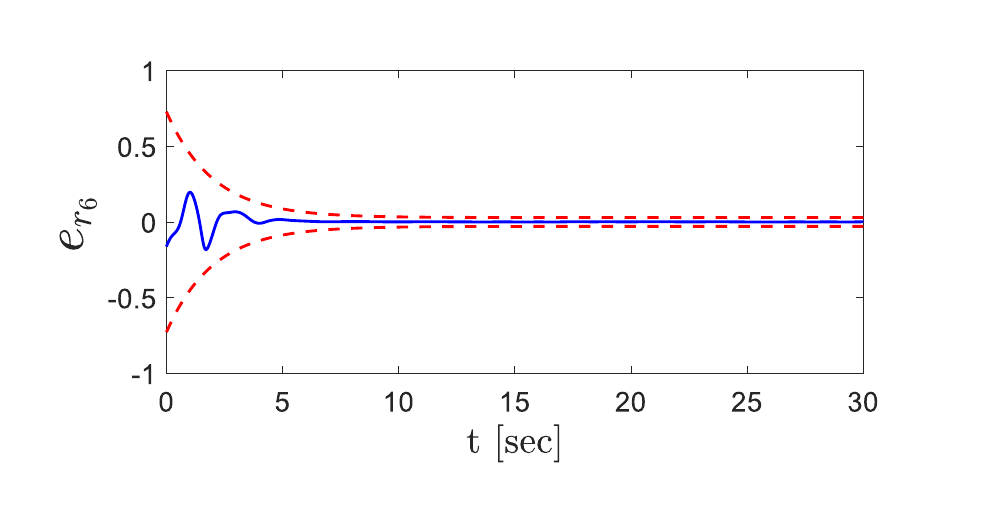}
	\end{subfigure}
	\caption{Evolution of the logarithmic ratio of the distance errors  $r_k, k\geq 3$. The magnified subplots provide the details of error evolution in the steady-state.}
	\label{fig:e_r}
\end{figure}

\begin{figure}[ht]
	\centering
	%	\flushleft
	\begin{subfigure}[t]{0.235\textwidth}
		\centering
		\includegraphics[width=\textwidth]{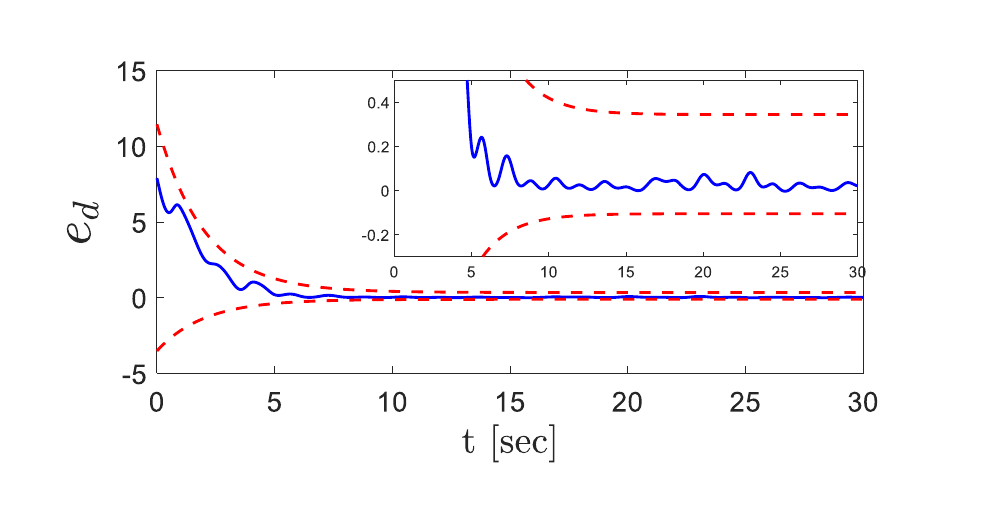}
	\end{subfigure}
	~
	\begin{subfigure}[t]{0.235\textwidth}
		\centering
		\includegraphics[width=\textwidth]{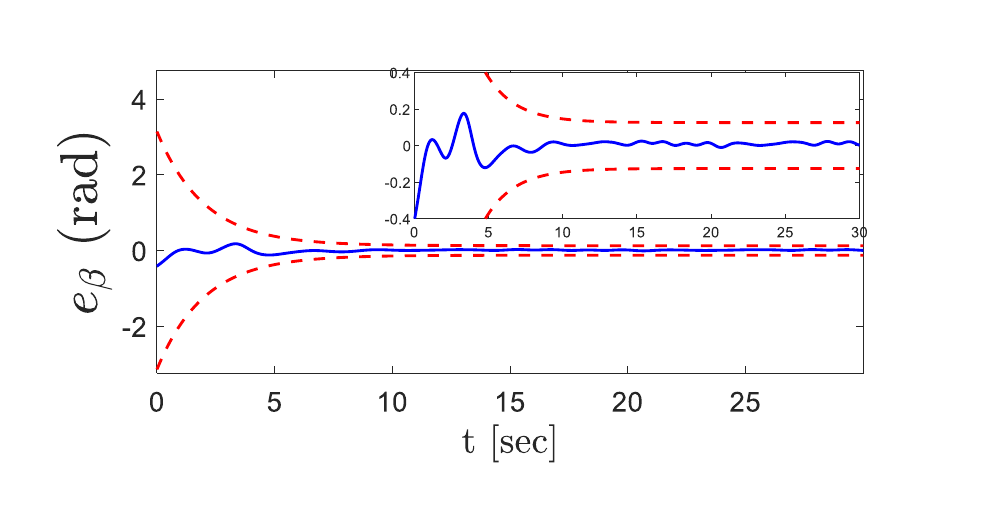}
	\end{subfigure}
	\caption{Evolution of agent 2's squared distance ($e_d$) and bearing angle ($e_{\beta}$) errors. The magnified subplots provide the details of error evolution in the steady-state.}
	\label{fig:e_d_beta}
\end{figure}

In this section, a simulation example of robust formation maneuvering with orientation and scaling control is presented to demonstrate the effectiveness of the proposed decentralized control protocols \eqref{eq:cont1}, \eqref{eq:cont2_extend}, and \eqref{eq:cont_k}\footnote{A short video demonstrating the following simulation results can be found at: https://youtu.be/jtsiU9DLp1k}.

 Consider a group of six agents modeled by \eqref{eq:singledyn} in a two-dimensional space. Suppose that the desired formation is an equilateral triangle composed of four equilateral sub-triangles (see Fig.\ref{fig:Maneu}), where its underlying sensing graph $\mathcal{G}$ satisfies Assumption \ref{assu:G} with the following directed edge set $\mathcal{E} =\{(2,1),(3,1),(3,2),(4,2),(4,3),(5,2),(5,4),(6,3),(6,4)\}$. Let the desired formation be characterized by the following sets of desired logarithmic ratio of the distances and edge-angles: $r^{\ast}_3 = r^{\ast}_4 = r^{\ast}_5 = r^{\ast}_6 = 0$, and $\alpha^{\ast}_{312}=\alpha^{\ast}_{634}=\pi/3$, $\alpha^{\ast}_{423}=\alpha^{\ast}_{524}=5 \pi / 3$, where $d_{31}^{\ast}=d_{32}^{\ast}=d_{42}^{\ast}=d_{43}^{\ast}=d_{52}^{\ast}=d_{53}^{\ast}=d_{63}^{\ast}=d_{64}^{\ast} = 1.875$. Moreover, assume that the local coordinate system of agent 2 (secondary leader), in which the formation orientation is defined, is aligned with the X-Y global coordinate system. Fig.\ref{fig:dist_beta_refs} shows the continuously differentiable time-varying reference signals $d_{21}^{\ast}(t)$ and $\beta^{\ast}(t)$ of agent 2 that adjust the formation scale and orientation as time goes on. Note that $d_{21}^{\ast}(t)$ and $\beta^{\ast}(t)$ are initially constant, and  $d_{21}^{\ast}(0\leq t \leq 16) = 1.875$  and $\beta^{\ast} (0\leq t \leq 13) = 0$. Without loss of generality, in the simulation we have assumed that the disturbance input to agent 1 (leader) is zero, $\delta_1(t) = 0$, and the leader follows a sinusoidal trajectory under the velocity control input of $u_L(t) = [1.25, \frac{\pi}{4} \cos(\frac{\pi}{6} t)]^T$. The external disturbances of agents  $\delta_{k} \coloneqq [\delta_{kx}, \delta_{ky}]^T$, $k=2,\ldots,6$, in the simulation are assumed to be: $\delta_{2x}(t) = \delta_{6y}(t) = 0.75\sin(4t+\frac{\pi}{5}) + 0.5\sin(2t+\frac{3\pi}{4})$, $\delta_{2y}(t) = \delta_{4y} (t) = \delta_{5x}(t) = 0.25 \cos(3t + \frac{\pi}{3}) + 0.75\sin(2t-\frac{\pi}{5})$, $\delta_{3x}(t) = 0.75\sin(t)$, $\delta_{3y}(t) = 0.25 \cos(t + \frac{\pi}{6}) + 0.25 \sin(2t+\frac{\pi}{4})$, $\delta_{4x}(t) = 0.5\cos(5t+\frac{\pi}{8}) + 0.5 \sin(t+\frac{\pi}{5})$, $\delta_{5y}(t) = 0.5 \cos(t)$, $\delta_{6x}(t) = 0.5\sin(2t+\frac{\pi}{4})$. Furthermore, the parameters of the performance functions \eqref{pf_rho} are considered as $l_{h}=0.5$, $\rho_{\infty,h}=0.04$, $h \in \{ \beta, r_k, \alpha_k  \}, k=\{3,\ldots,6\}$, and $l_{d}=0.5$, $\rho_{\infty,d}=0.03$. Moreover, the positive constants $\underline{b}_{h}, \bar{b}_{h}$, $h \in \{d,\beta, r_k, \alpha_k  \}, k=\{3,\ldots,6\}$ of the performance bounds are selected according to the guidelines in Sections \ref{sec:bounds_select} and \ref{subsec:orientation}.

Considering the aforementioned setting as well as a set of arbitrary initial positions for the agents, the results are summarized in Fig.\ref{fig:Maneu}, which depicts consecutive snapshots of the agents' trajectories towards the desired formation as the leader follows its reference trajectory. Note that agent 2 starts tracking a time-varying bearing angle with respect to agent 1 from $t>13$ (see Fig.\ref{fig:dist_beta_refs}) such that it follows the angle of the leader's velocity direction with an offset of $-\pi/6$ radians. In addition, the evolution of $d_{21}^{\ast}(t)$ during $16 < t < 26$ allows agents to pass through a narrow passage without colliding with obstacles when following the leader and maintaining the desired shape. The evolution of the edge-angle \eqref{eq:alpha_error} and the logarithmic ratio of the distances \eqref{eq:r_error} errors for agents $3 \leq k \leq 6$ are depicted in Fig.\ref{fig:e_alpha} and Fig.\ref{fig:e_r}, respectively, where the dashed lines indicate the user-defined performance bounds. Moreover, Fig.\ref{fig:e_d_beta} shows the evolution of bearing angle error \eqref{eq:e_beta} as well as the squared distance error \eqref{eq:eij} of agent 2. Notice that, the formation errors remain within the pre-defined performance bounds for all time. Hence, the results indicate that the proposed formation control scheme is capable of handling the problem of coordinate-free (stationary or maneuvering) formation control with adjustable scaling and orientation as well as global shape convergence under prescribed performance specifications that further introduce robustness to external disturbances.

%################################################################
\section{Conclusions}
\label{Sec:concu}

In this paper, we proposed a novel 2-D directed formation control approach with (almost) global convergence using bipolar coordinates for desired shapes modeled by acyclic triangulated directed graphs (a class of acyclic minimally persistent graphs). Bipolar coordinates were used to characterize the desired formation to avoid undesired equilibria in 2-D coordinate-free directed formations. Then the prescribed performance control method was adopted for designing the formation control laws to introduce robustness against external disturbances/model uncertainties as well as ensuring user-defined transient and steady-state performance guarantees. We further showed that the proposed approach is capable of handling formation maneuvering with time-varying reference velocities along with scaling and orientation adjustment. Moreover, it was argued that the control approach can be easily implemented in arbitrarily oriented local coordinate frames of the (follower) agents by using onboard vision sensors, which are favorable for practical applications. Future research efforts will be devoted to extending the results for 3-D formations as well as dealing with inter-agent collision avoidance of non-point agents. Moreover, handling the problem of connectivity maintenance for the proposed control methodology is also an important direction of research for the future. Finally, considering more complex agent dynamics (i.e., higher-order and nonlinear) and also taking into account sensing noise in the feedback control are among other interesting topics for future work.

%################################################################
\appendices

\section{Proof of Lemma \ref{lem:bipolar_basis_global}}
\label{appen:proof_lem_bipolar_basis}
From \eqref{eq:bipolar_cord_in_virtual} recall that the position of node $k$ w.r.t.\ $\left\lbrace C_k \right\rbrace$ is $p_k^{[C_k]} = [x_k^{[C_k]}, y_k^{[C_k]}]^T$. Let $\widehat{x}_k^{[C_k]}, \widehat{y}_k^{[C_k]} \in \mathbb{R}^2$ be the (unit) orthogonal bases of $\left\lbrace C_k \right\rbrace$ that are expressed in $\left\lbrace C_k \right\rbrace$, hence:
\begin{equation} \label{eq:pk_in_ck}
	p_k^{[C_k]} = x_k^{[C_k]} \, \widehat{x}_k^{[C_k]} + y_k^{[C_k]} \, \widehat{y}_k^{[C_k]},
\end{equation}
where, $x_k^{[C_k]}$ and $y_k^{[C_k]}$ are given in \eqref{eq:bipolar_cord_in_virtual}. It is known that for the bipolar coordinates (that is an orthogonal curvilinear coordinate system) associated with $\left\lbrace C_k \right\rbrace$, the two vectors $\widehat{\alpha}_{k}^{[C_k]}$, $\widehat{r}_k^{[C_k]}$ form a local basis at any nonsingular point $p_k^{[C_k]}$, where the following hold \cite{happel2012low}:
\begin{equation} \label{eq:curvili_hat_relation}
	\dfrac{\partial p_k^{[C_k]}}{\partial \alpha_{kij}} = q_{\alpha_{k}} \widehat{\alpha}_{k}^{[C_k]}, \; \text{and} \; \dfrac{\partial p_k^{[C_k]}}{\partial r_k} = q_{r_{k}} \widehat{r}_{k}^{[C_k]},
\end{equation}
in which
\begin{equation} \label{eq:scalingfactors}
	q_{\alpha_{k}} = q_{r_{k}} = \dfrac{c_k}{\cosh r_k - \cos \alpha_{kij}},
\end{equation}
are the scaling (metrical) factors \cite{weisstein2002crc, happel2012low} and $c_k$ was defined in \eqref{eq:bipolar_cord_in_virtual}. From \eqref{eq:bipolar_cord_in_virtual}, \eqref{eq:pk_in_ck}, \eqref{eq:curvili_hat_relation}, and \eqref{eq:scalingfactors} one can obtain:
\begin{subequations}\label{eq:bipolar_basis_in_C}
	\begin{align}
		\widehat{\alpha}_{k}^{[C_k]} &= f_1(r_k, \alpha_{kij}) \, \widehat{x}_k^{[C_k]} + f_2(r_k, \alpha_{kij}) \, \widehat{y}_k^{[C_k]}, \\
		\widehat{r}_{k}^{[C_k]} &= -f_2(r_k, \alpha_{kij}) \, \widehat{x}_k^{[C_k]} + f_1(r_k, \alpha_{kij}) \, \widehat{y}_k^{[C_k]}, 
	\end{align}
\end{subequations}
where $f_1, f_2$ are given in \eqref{eq:B1_B2}. Notice that:
\begin{subequations}\label{eq:C_basis_in_global}
	\begin{align}
		\widehat{x}_{k}^{[C_k]} &= -\dfrac{{p}_{ji}}{\| {p}_{ji} \|} = - z_{ji},\\
		\widehat{y}_{k}^{[C_k]} &= - J\dfrac{{p}_{ji}}{\| {p}_{ji} \|} = - J z_{ji} = J^T z_{ji},
	\end{align}
\end{subequations}
which provide the basis of $\left\lbrace C_k \right\rbrace$ in the global coordinate frame. Finally, \eqref{eq:bipolar_basis_in_C} and \eqref{eq:C_basis_in_global} yield \eqref{eq:bipolar_basis_in_globalframe}.

\section{Proof of Theorem \ref{th:agent2}} \label{appen:th1}
\textit{Phase I.} First notice that agent 2's formation error dynamics can be obtained invoking \eqref{eq:singledyn}, \eqref{eq:eij}, \eqref{eq:cont1} as follows:
\begin{align}
	\dot{e}_d = 2{p}_{21}^{T} \dot{p}_{21} - 2d_{21}^{\ast} \dot{d}_{21}^{\ast}  &= 2 {p}_{21}^{T} (u_L - u_{2} + \delta_{21}) - 2 d_{21}^{\ast} \dot{d}_{21}^{\ast} \nonumber \\
	&= 2{p}_{21}^{T}(\Lambda_2 - u_2) - \Gamma_d, \label{eq: dot_e21}
\end{align}
where $\delta_{21} \coloneqq \delta_1 - \delta_2 \in \mathbb{R}^2$, $\Lambda_2 \coloneqq u_L + \delta_{21} \in \mathbb{R}^2$, $\Gamma_d \coloneqq 2 d_{21}^{\ast} \dot{d}_{21}^{\ast} \in \mathbb{R}$ are uniformly bounded signals ($\delta_{21}, \Lambda_2, \Gamma_d \in \mathcal{L}_{\infty}$) by assumption. Now differentiating $\tilde{e}_d(t)$ in \eqref{modu_err} and employing \eqref{eq: dot_e21}, \eqref{eq:cont2}, yields:
\begin{align} 
	\dot{\tilde{e}}_d &\coloneqq E_{\tilde{e}_d}(t,\tilde{e}_d) =\rho_d^{-1}(t) \left(\dot{e}_d-\tilde{e}_d \dot{\rho}_d(t)\right) \nonumber \\ 
	=& \rho_d^{-1} \left( 2{p}_{21}^{T}\Lambda_2 - 2 \xi_d \sigma_d \|{p}_{21}\|^2 - \Gamma_d - \tilde{e}_d \dot{\rho}_d \right). \label{eq:e_d_modu_dyn}
\end{align}
Let us also define the open set $\Omega_{\tilde{e}_d}$ as: $\Omega_{\tilde{e}_d} \coloneqq (-\underline{b}_{d}, \bar{b}_{d})$.
Note that, $\Omega_{\tilde{e}_d}$ is nonempty and open by construction. Moreover, followed by the discussion in Section \ref{sec:bounds_select}, agent 2 can always initially select  $\underline{b}_{d}, \bar{b}_{d}>0$ to ensure $\tilde{e}_d(0) \in \Omega_{\tilde{e}_d}$. Additionally, $E_{\tilde{e}_d}(t,\tilde{e}_d)$ is continuous on $t$ and locally Lipschitz on $\tilde{e}_d$ over the set $\Omega_{\tilde{e}_d}$. Therefore, the hypotheses of Theorem 54 in \cite[p.~476]{sontag1998mathematical} hold and the existence and uniqueness of a maximal solution $\tilde{e}_d(t)$ of \eqref{eq:e_d_modu_dyn} for a time interval $[0, \tau_{\mathrm{2,max}})$ such that $\tilde{e}_d(t) \in \Omega_{\tilde{e}_d}, \forall t \in [0, \tau_{\mathrm{2,max}})$ is guaranteed. Based on this, we can further infer that ${e}_{d}(t)$ is  bounded as in \eqref{eq:e_h_bound} for all $t \in [0, \tau_{\mathrm{2,max}})$.

\textit{Phase II.} Owing to $\tilde{e}_d(t) \in \Omega_{\tilde{e}_d}, \forall t \in [0, \tau_{\mathrm{2,max}})$, the error $\sigma_d$, as defined in \eqref{eq:mappings}, is well-defined for all $t \in [0, \tau_{\mathrm{2,max}})$. Therefore, consider the following positive definite and radially unbounded Lyapunov function candidate: $V_2 = (1/4) \sigma_d^2$. Taking the time derivative of $V_2$, invoking \eqref{eq: derivat_map_k}, \eqref{eq: dot_e21}, \eqref{eq:cont2}, and the positivity of $\xi_d$, we get:
\begin{align}
	\dot{V}_2 &= - \xi_d^2 \sigma_d^2 \|p_{21}\|^2 + \xi_d \sigma_d p_{21}^T \Lambda_2 - \xi_d \sigma_d \dfrac{1}{2}(\Gamma_d + \tilde{e}_d \dot{\rho}_d) \nonumber \\
	&\leq - \xi_d^2 \sigma_d^2 \|p_{21}\|^2 + \xi_d |\sigma_d| \|p_{21}\| \|\Lambda_2\| + \xi_d |\sigma_d| |\Psi_2|, \label{V2_deriv}
\end{align}
where $\Psi_2 \coloneqq 0.5(\Gamma_d + \tilde{e}_d \dot{\rho}_d) \in \mathbb{R}$, which is bounded for all $t \in [0, \tau_{\mathrm{2,max}})$	owing to the boundedness of $\dot{\rho_{d}}(t), \Gamma_d(t)$ and $\tilde{e}_d(t)$ for $\forall t \in [0, \tau_{\mathrm{2,max}})$ (as it was shown in \textit{Phase I}). Let $0<\theta_{2}<1$ be a constant; thus adding and subtracting $\theta_2 \xi_{d}^2 \sigma_d^2  \|p_{21}\|^2$ to the right-hand side of \eqref{V2_deriv} yields:
\begin{equation} \label{eq:V2_deriv_final}
	\begin{split}
		\dot{V}_2 \leq& - (1 - \theta_2) \xi_d^2 \sigma_d^2 \|p_{21}\|^2  \\ 
		&- \xi_d |\sigma_d| \left(\theta_2 \xi_{d} |\sigma_d|  \|p_{21}\|^2 - \|p_{21}\| \|\Lambda_2\| - |\Psi_2|\right) \\
		\leq& - (1 - \theta_2) \xi_d^2 \sigma_d^2 \|p_{21}\|^2, \\
		\forall& \, |\sigma_d| \geq \frac{\|\Lambda_2\| \|p_{21}\| + |\Psi_2|}{\theta_{2} \xi_{d} \|p_{21}\|^2},  \quad \forall t \in [0, \tau_{\mathrm{2,max}}).
	\end{split}
\end{equation} 
Recall that $\Lambda_2(t), \Psi_2(t), \theta_{2} \in \mathcal{L}_{\infty}, \forall t \in [0, \tau_{\mathrm{2,max}})$. Notice that $\xi_d$ is lower bounded by a positive constant. In addition, since $\tilde{e}_d(t) \in \Omega_{\tilde{e}_d} = (-\underline{b}_{d}, \bar{b}_{d}), \forall t \in [0, \tau_{\mathrm{2,max}})$, followed by \eqref{modu_err}, \eqref{eq:eij}, \eqref{eq:e_h_bound}, and Section \ref{sec:bounds_select}, $\|p_{21}(t)\|^2 >  \inf_{t \in [0, \tau_{\mathrm{2,max}}) } \left( (d_{21}^{\ast}(t))^2 - \underline{b}_d \rho_d(t) \right)  > 0 $  and $\|p_{21}(t)\|^2 < \sup_{t \in [0, \tau_{\mathrm{2,max}}) } \left( (d_{21}^{\ast}(t))^2 + \bar{b}_d \rho_d(t) \right)$ for all $t \in [0, \tau_{\mathrm{2,max}})$. Therefore, \eqref{eq:V2_deriv_final} indicates that $\sigma_d(t)$ is Uniformly Ultimately Bounded (UUB) \cite{khalil2002noninear}, and one can show that there exists an ultimate bound $\bar{\sigma}_d $
%$\bar{\sigma}_d = \max\left\lbrace  (\|\Lambda_2\| \|p_{21}\| + |\Psi_2|) / (\theta_{2} \xi_{d} \|p_{21}\|^2) \right\rbrace  < \infty$ 
independent of $\tau_{\text{2,max}}$ 
such that $|\sigma_d(t)| \leq \bar{\sigma}_d$ for $\forall t \in [0, \tau_{\mathrm{2,max}})$.

\textit{Phase III.}  Owing to the properties of $T_{d}(\tilde{e}_d)$ in \eqref{eq:mappings}, we have: $-\underline{b}_d < T_{d}^{-1}(\sigma_d) = \tilde{e}_{d} < \bar{b}_d$. Furthermore, since $T_{d}^{-1}(\sigma_d)$ is strictly increasing and $|\sigma_d(t)|\leq \bar{\sigma}_d$ there exist $-\underline{b}_d^{\ast}(\bar{\sigma}_d), \bar{b}_d^{\ast}(\bar{\sigma}_d)$ such that:
\begin{equation} \label{eq:eta_hat2_in_bound}
	-\underline{b}_d < -\underline{b}_d^{\ast}(\bar\sigma_d) \leq \tilde{e}_{d} \leq \bar{b}_d^{\ast}(\bar\sigma_d)  < \bar{b}_d.
\end{equation}
As a result $\tilde{e}_{d}(t) \in \Omega_{\tilde{e}_{d}}^{\prime}, \forall t \in [0, \tau_{\mathrm{2,max}})$ where $\Omega_{\tilde{e}_{d}}^{\prime} = [-\underline{b}_d^{\ast}, \bar{b}_d^{\ast}]$ is a nonempty compact subset of $\Omega_{\tilde{e}_{d}}$. Hence, assuming a finite $\tau_{\mathrm{2,max}} < \infty$ and since $\Omega_{\tilde{e}_{d}}^{\prime} \subset \Omega_{\tilde{e}_{d}}$, Proposition C.3.6 in \cite[p.~481]{sontag1998mathematical} dictates the existence of a time instant $t^{\prime} \in [0,\tau_{\mathrm{2,max}})$ such that $\tilde{e}_{d}(t^{\prime}) \notin \Omega_{\tilde{e}_{d}}^{\prime}$, which is a contradiction. Therefore, $\tau_{\mathrm{2,max}} = \infty$. Thus, all closed loop signals remain bounded and moreover $\tilde{e}_{d}(t) \in \Omega_{\tilde{e}_{d}}^{\prime} \subset \Omega_{\tilde{e}_{d}}, \forall t \geq 0$. Multiplying \eqref{eq:eta_hat2_in_bound} by $\rho_{d}(t)$ results in: $-\underline{b}_{d}\rho_{d}(t) < -\underline{b}_d^{\ast} \rho_{d}(t) \leq e_{d}(t) \leq \bar{b}_d^{\ast} \rho_{d}(t) < \bar{b}_{d} \rho_{d}(t)$ for all $t \geq 0$, which further ensures  \eqref{eq:e_h_bound} and thus $\|p_{21}(t)\| > 0$, for all $t\geq0$, due to selection of $\underline{b}_d, \bar{b}_d$ according to Section \ref{sec:bounds_select}. 

\section{Proof of Theorem \ref{th:agent_k}} \label{appen:th2}
First note that, using \eqref{eq:singledyn}, \eqref{eq:log_ratio}, \eqref{eq:r_error}, and \eqref{eq:alpha_error}, the formation error dynamics of agent $k\geq3$ is given by:
\begin{subequations} \label{eq:err_dyn}
	\begin{align}
		\dot{e}_{r_k} &= \dfrac{p_{ki}^T \dot{p}_{ki}}{\|p_{ki}\|^2} - \dfrac{p_{kj}^T \dot{p}_{kj}}{\|p_{kj}\|^2}  \nonumber \\ 
		&=\dfrac{z_{ki}^T}{\|{p}_{ki}\|} ( u_i - u_k + \delta_{ki}) - \dfrac{z_{kj}^T}{\|{p}_{kj}\|} (u_j - u_k +\delta_{kj}) \nonumber \\
		&=\dfrac{z_{ki}^T}{\|{p}_{ki}\|} (\Lambda_{ki} - u_k) - \dfrac{z_{kj}^T}{\|{p}_{kj}\|} (\Lambda_{kj} - u_k), \label{eq: dot_r_k}
	\end{align}
	\begin{align}
		\dot{e}_{\alpha_k} &=  \dfrac{z_{ki}^T}{\|{p}_{ki}\|} J (u_i- u_k +\delta_{ki}) 
		- \dfrac{z_{kj}^T}{\|{p}_{kj}\|} J (u_j - u_k + \delta_{kj}) \nonumber \\
		&=  \dfrac{z_{ki}^T}{\|{p}_{ki}\|} J (\Lambda_{ki}- u_k) 
		- \dfrac{z_{kj}^T}{\|{p}_{kj}\|} J (\Lambda_{kj} - u_k), \label{eq: dot_alpha_k}
	\end{align}	
\end{subequations}
where $(k,i), (k,j) \in \mathcal{E} \setminus \{(2,1)\} , i<j<k$, and $\delta_{ki} \coloneqq \delta_i - \delta_k, \delta_{kj} \coloneqq \delta_j - \delta_k$, $\Lambda_{ki} \coloneqq u_i + \delta_{ki}$, $\Lambda_{kj} \coloneqq u_j + \delta_{kj}$. Note that  \eqref{eq: dot_alpha_k} is obtained by using the arc length formula (see Appendix \ref{appen:alpha_tild_dot_calc}). Define $e_k \coloneqq [{e}_{r_k}, {e}_{\alpha_k}]^T \in \mathbb{R}^2, k = 3,\ldots ,n$, as the stacked formation errors for agent $k$. Based on \eqref{eq:err_dyn} we have:
\begin{equation} 
	\dot{e}_k = H_k \Lambda_k + G_k u_k, \quad k = 3\ldots,n, \label{eq:compact_e_k_dot}
\end{equation}
where $\Lambda_k \coloneqq [\Lambda_{ki}^\top, \Lambda_{kj}^\top]^\top \in \mathbb{R}^{4}$, and $H_k \in \mathbb{R}^{2 \times 4}, G_k \in \mathbb{R}^{2 \times 2}$ are as follows:
\begin{equation*}
	H_k \coloneqq
	\begin{bmatrix*}[l]
		\frac{z_{ki}^T}{\|p_{ki}\|} & -\frac{z_{kj}^T}{\|p_{kj}\|} \\[5pt]
		\frac{z_{ki}^T}{\|p_{ki}\|}J & -\frac{z_{kj}^T}{\|p_{kj}\|}J
	\end{bmatrix*}, \;
	G_k \coloneqq
	\begin{bmatrix*}[c]
		\frac{z_{kj}^T}{\|p_{kj}\|} - \frac{z_{ki}^T}{\|p_{ki}\|} \\[5pt]
		\left(\frac{z_{kj}^T}{\|p_{kj}\|} - \frac{z_{ki}^T}{\|p_{ki}\|} \right) J
	\end{bmatrix*}.
\end{equation*}
Moreover, defining the stacked transferred formation errors as $\sigma_k \coloneqq [\sigma_{r_k}, \sigma_{\alpha_k}]^T$, $k = 3,\ldots,n$, and employing \eqref{eq: derivat_map_k} gives: 
\begin{equation} \label{eq:compact_sigma_k_dot}
	\dot{\sigma}_k =  \xi_k (\dot{e}_k - \dot{\rho}_k \tilde{e}_k ),
\end{equation}
where $\xi_k \coloneqq \mathrm{diag}(\xi_{r_k}, \xi_{\alpha_{k}}) \in \mathbb{R}^{2 \times 2}$, $\rho_k \coloneqq \mathrm{diag}(\rho_{r_k}, \rho_{\alpha_{k}}) \in \mathbb{R}^{2 \times 2}$, and $\tilde{e}_k \coloneqq [\tilde{e}_{r_k}, \tilde{e}_{\alpha_k}]^T = \rho_k^{-1} e_k$. Finally notice that, the control law \eqref{eq:cont_k} can be re-written as follows:
\begin{equation} \label{eq:compact_u_k}
	u_k = -B_k \xi_k \sigma_k, \quad k = 3\ldots,n,
\end{equation}
where $B_k \coloneqq [\widehat{r}_k \,|\, \widehat{\alpha}_k] \in \mathbb{R}^{2 \times 2}, k = 3\ldots,n$, are matrices whose columns are the orthogonal bipolar basis associated with agent $k\geq3$. In the following, we shall first establish the results for agent 3 and then extend the proof for all $3<k\leq n$ by induction. Similarly to the proof of Theorem \ref{th:agent2}, we will proceed in three phases.

\textit{Phase I.} Differentiating $\tilde{e}_{r_3}$ and $\tilde{e}_{\alpha_{3}}$, gives:
\begin{subequations}
	\begin{align} 
		\dot{\tilde{e}}_{r_3} &\coloneqq E_{\tilde{e}_{r_3}} (t,\tilde{e}_{r_3}) = \rho_{r_3}^{-1}(t) \left(\dot{e}_{r_3}-\tilde{e}_{r_3} \dot{\rho}_{r_3}(t)\right), \label{eq:e_r_3_modu_dyn} \\
		\dot{\tilde{e}}_{\alpha_3} 	&\coloneqq E_{\tilde{e}_{\alpha_3}} (t,\tilde{e}_{\alpha_3}) = \rho_{\alpha_3}^{-1}(t) \left(\dot{e}_{\alpha_3}-\tilde{e}_{\alpha_3} \dot{\rho}_{\alpha_3}(t)\right)
	\end{align}
\end{subequations}
Define $E_{\tilde{e}_{3}}(t, \tilde{e}_3) \coloneqq [E_{\tilde{e}_{r_3}} (t,\tilde{e}_{r_3}), E_{\tilde{e}_{\alpha_3}}(t,\tilde{e}_{\alpha_3})]^T$. Using \eqref{eq:compact_e_k_dot} and \eqref{eq:compact_u_k}, the closed-loop dynamical system of $\tilde{e}_3 = [\tilde{e}_{r_3}, \tilde{e}_{\alpha_3}]^T = \rho_3^{-1} e_3$ with $\rho_3 = \mathrm{diag}(\rho_{r_3}, \rho_{\alpha_{3}})$ may be written in compact form as:
\begin{align}
	\dot{\tilde{e}}_3 = E_{\tilde{e}_{3}}&(t, \tilde{e}_3) = \rho_3^{-1}(t) (\dot{e}_3 - \dot{\rho}_3(t) \tilde{e}_3) \nonumber \\
	=& \rho_3^{-1}(t) \left( H_3 \Lambda_3 - G_3 B_3 \xi_3 \sigma_3 - \dot{\rho}_3(t) \tilde{e}_3\right). \label{eq:tilde_e_3_dyn}
\end{align}
Let us also define the open set: $\Omega_{\tilde{e}_3} \coloneqq \Omega_{\tilde{e}_{r_3}} \times \Omega_{\tilde{e}_{\alpha_{3}}}$, where $\Omega_{\tilde{e}_{r_3}} \coloneqq (-\underline{b}_{r_3}, \bar{b}_{r_3})$, and $\Omega_{\tilde{e}_{\alpha_3}} \coloneqq (-\underline{b}_{\alpha_3}, \bar{b}_{\alpha_3})$. Note that $\Omega_{\tilde{e}_3}$ is nonempty and open by construction. Followed by the discussion in Section \ref{sec:bounds_select}, agent 3 can initially select  $\underline{b}_{r_3}, \bar{b}_{r_3}, \underline{b}_{\alpha_3}, \bar{b}_{\alpha_3}>0$ to ensure $\tilde{e}_3(0) \in \Omega_{\tilde{e}_3}$. Since $E_{\tilde{e}_3}(t,\tilde{e}_3)$ is continuous on $t$ and locally Lipschitz on $\tilde{e}_3$ over the set $\Omega_{\tilde{e}_3}$, the hypotheses of Theorem 54 in \cite[p.~476]{sontag1998mathematical} dictates existence and uniqueness of a maximal solution $\tilde{e}_3(t)$ of \eqref{eq:tilde_e_3_dyn} for a time interval $[0, \tau_{\mathrm{3,max}})$ where $\tilde{e}_3(t) \in \Omega_{\tilde{e}_3}, \forall t \in [0, \tau_{\mathrm{3,max}})$ is guaranteed. This further ensures that $\tilde{e}_{r_3}(t)$ and $\tilde{e}_{\alpha_{3}}(t)$ are bounded as in \eqref{eq:e_h_bound} for all $t \in [0, \tau_{\mathrm{3,max}})$.

\textit{Phase II.} Owing to $\tilde{e}_3(t) \in \Omega_{\tilde{e}_3}, \forall t \in [0, \tau_{\mathrm{3,max}})$, the stacked transformed errors $\sigma_3 = [\sigma_{r_3}, \sigma_{\alpha_3}]^T$, where $\sigma_{r_3}, \sigma_{\alpha_3}$ are defined in \eqref{eq:mappings}, are well-defined for all $t \in [0, \tau_{\mathrm{3,max}})$. Therefore, consider the following positive definite and radially unbounded Lyapunov function candidate $V_3 = (1/2) \sigma_3^T \sigma_3$. Differentiating $V_3$ with respect to time, using \eqref{eq:compact_e_k_dot}, \eqref{eq:compact_sigma_k_dot}, and \eqref{eq:compact_u_k}, gives:
\begin{align}
	\dot{V}_3 =& - \sigma_3^T \xi_3 \left(G_3 B_3\right) \xi_3 \sigma_3 + \sigma_3^T \xi_3 H_3 \Lambda_3 - \sigma_3^T \xi_3 \dot{\rho}_3 \tilde{e}_3,  \label{V3_deriv} \\ 
	\leq& - \underline{m}_3 \|\sigma_3^T \xi_3\|^2 + \|\sigma_3^T \xi_3\| \|H_3\| \|\Lambda_3\| + \|\sigma_3^T \xi_3\| \|\Psi_3\|, \nonumber 
\end{align}
where $\underline{m}_3>0$ is a positive constant related to $M_3 \coloneqq G_3 B_3$ (see Appendix \ref{appen:P.D. G_k B_k} for details), and $\Psi_3 \coloneqq \dot{\rho}_3 \tilde{e}_3 \in \mathbb{R}^2$, which is bounded for $\forall t \in [0, \tau_{\mathrm{3,max}})$	owing to the boundedness of $\dot{\rho_{3}}(t)$ for all $t\geq 0$ and the boundedness of $\tilde{e}_3(t)$ for $\forall t \in [0, \tau_{\mathrm{3,max}})$ (as it is shown in \textit{Phase I}). Moreover, note that due to the boundedness of $\delta_{31}, \delta_{32}, u_1 = u_L(t) \in \mathcal{L}_{\infty}$ as well as boundedness of $u_2(t) \in \mathcal{L}_{\infty}$ (owing to Theorem \ref{th:agent2}), we have that $\Lambda_{31}, \Lambda_{32} \in \mathcal{L}_{\infty}$, which leads to the boundedness of $\Lambda_{3}\in \mathcal{L}_{\infty}$. Let $0<\theta_{3}<\underline{m}_3$ be a constant, adding and subtracting $\theta_3 \|\sigma_3^T \xi_{3} \|^2$ to the right-hand side of \eqref{V3_deriv}, and invoking diagonality and positive definiteness of $\xi_3$, yields:
\begin{align}
	\dot{V}_3 \leq& - (\underline{m}_3 - \theta_{3}) \|\sigma_3^T \xi_3\|^2 \nonumber  \\
	& - \|\sigma_3^T \xi_3\| \left(\theta_{3} \|\sigma_3^T \xi_3\| - \|H_3\| \|\Lambda_3\| - \|\Psi_3\|\right) \nonumber  \\
	\leq& - (\underline{m}_3 - \theta_{3}) \lambda_{\min}(\xi_3^2) \|\sigma_3\|^2, \label{eq:Lyap_3_deriv_final}  \\
	\forall& \, \|\sigma_3^T \xi_3 \| \geq \frac{\|H_3\| \|\Lambda_3\|  + \|\Psi_3\|}{\theta_{3}},  \quad \forall t \in [0, \tau_{\mathrm{3,max}}), \nonumber
\end{align} 
where $\lambda_{\min}(\xi_3^2)$ is the minimum eigenvalue of the diagonal positive definite matrix $\xi_3^2 \in \mathbb{R}^{2 \times 2}$. Note that, $\tilde{e}_3(t) \in \Omega_{\tilde{e}_3} = (-\underline{b}_{r_3}, \bar{b}_{r_3}) \times (-\underline{b}_{\alpha_3}, \bar{b}_{\alpha_3}), \forall t \in [0, \tau_{\mathrm{3,max}})$, hence followed by \eqref{eq:log_ratio}, \eqref{eq:r_error}, \eqref{modu_err}, \eqref{eq:e_h_bound}, and Section \ref{sec:bounds_select} for the selection of $\underline{b}_{r_3}, \bar{b}_{r_3}$, we can infer that $\|p_{31}(t)\|, \|p_{32}(t)\|$ are bounded away from zero for $\forall t \in [0, \tau_{\mathrm{3,max}})$. Moreover, since $J , z_{31}, z_{32} \in \mathcal{L}_{\infty}$, the elements of matrix $H_3$ are all bounded for $\forall t \in [0,\tau_{\mathrm{3,max}})$, thus $\|H_3\| \in \mathcal{L}_{\infty}, \forall t \in [0,\tau_{\mathrm{3,max}})$. Finally, as $\|H_3(t)\|, \|\Lambda_3(t)\|, \|\Psi_3(t)\|, \theta_{3} \in \mathcal{L}_{\infty}, \forall t \in [0, \tau_{\mathrm{3,max}})$, and $\xi_3$ is a diagonal positive definite matrix, \eqref{eq:Lyap_3_deriv_final} implies that $\sigma_3$ is uniformly ultimately bounded \cite{khalil2002noninear}. Therefore, one can show that there exists an ultimate bound $\bar{\sigma}_3$ independent of $\tau_{\text{3,max}}$ such that $\|\sigma_3(t)\| \leq \bar{\sigma}_3$ for $\forall t \in [0, \tau_{\mathrm{3,max}})$. 

\textit{Phase III.}  Owing to $\|\sigma_3(t)\| \leq \bar{\sigma}_3$ we have $|\sigma_{r_3}(t)| \leq \bar{\sigma}_3$ and $|\sigma_{\alpha_3}(t)| \leq \bar{\sigma}_3$. Similarly to Phase III in the proof of Theorem \ref{th:agent2}, due to properties of $T_{h}(\tilde{e}_h)$ in \eqref{eq:mappings} and its inverse, there exist $-\underline{b}_{r_3}^{\ast}(\bar{\sigma}_3), \bar{b}_{r_3}^{\ast}(\bar{\sigma}_3), -\underline{b}_{\alpha_3}^{\ast}(\bar{\sigma}_3), \bar{b}_{\alpha_3}^{\ast}(\bar{\sigma}_3)$ such that:
\begin{subequations}\label{eq:eta_hat3_in_bound}
	\begin{align} 
		-\underline{b}_{r_3} < -\underline{b}_{r_3}^{\ast}(\bar\sigma_{3}) \leq &\tilde{e}_{r_3} \leq \bar{b}_{r_3}^{\ast}(\bar\sigma_{3})  < \bar{b}_{r_3}, \label{eq:a} \\
		-\underline{b}_{\alpha_3} < -\underline{b}_{\alpha_3}^{\ast}(\bar\sigma_{3}) \leq &\tilde{e}_{\alpha_3} \leq \bar{b}_{\alpha_3}^{\ast}(\bar\sigma_{3})  < \bar{b}_{\alpha_3}. \label{eq:b}
	\end{align}
\end{subequations}
As a result $\tilde{e}_{3}(t) \in \Omega_{\tilde{e}_{3}}^{\prime} \coloneqq \Omega_{\tilde{e}_{r_3}}^{\prime} \times \Omega_{\tilde{e}_{\alpha_{3}}}^{\prime}, \forall t \in [0, \tau_{\mathrm{3,max}})$ where $\Omega_{\tilde{e}_{r_3}}^{\prime} = [-\underline{b}_{r_3}^{\ast}, \bar{b}_{r_3}^{\ast}]$, $\Omega_{\tilde{e}_{\alpha_3}}^{\prime} = [-\underline{b}_{\alpha_3}^{\ast}, \bar{b}_{\alpha_3}^{\ast}]$ are nonempty compact subset of $\Omega_{\tilde{e}_{r_3}}$ and $\Omega_{\tilde{e}_{\alpha_3}}$, respectively. Hence, assuming a finite $\tau_{\mathrm{3,max}} < \infty$, since $\Omega_{\tilde{e}_{3}}^{\prime} \subset \Omega_{\tilde{e}_{3}}$, Proposition C.3.6 in \cite[p.~481]{sontag1998mathematical} leads to the existence of a time instant $t^{\prime} \in [0,\tau_{\mathrm{3,max}})$ such that $\tilde{e}_{3}(t^{\prime}) \notin \Omega_{\tilde{e}_{3}}^{\prime}$, which is a contradiction. Therefore, $\tau_{\mathrm{3,max}} = \infty$. Thus, all closed loop signals remain bounded, and $\tilde{e}_{3}(t) \in \Omega_{\tilde{e}_{3}}^{\prime} \subset \Omega_{\tilde{e}_{3}}, \forall t \geq 0$. Multiplying \eqref{eq:a} and \eqref{eq:b} by $\rho_{r_3}(t)$ and $\rho_{\alpha_3}(t)$, respectively, gives: $-\underline{b}_{r_3}\rho_{r_3}(t) < -\underline{b}_{r_3}^{\ast} \rho_{r_3}(t) \leq e_{r_3}(t) \leq \bar{b}_{r_3}^{\ast} \rho_{r_3}(t) < \bar{b}_{r_3} \rho_{r_3}(t)$ and $-\underline{b}_{\alpha_3}\rho_{\alpha_3}(t) < -\underline{b}_{\alpha_3}^{\ast} \rho_{\alpha_3}(t) \leq e_{\alpha_3}(t) \leq \bar{b}_{\alpha_3}^{\ast} \rho_{\alpha_3}(t) < \bar{b}_{\alpha_3} \rho_{\alpha_3}(t)$ for $t \geq 0$, which ensure \eqref{eq:e_h_bound} for $e_{r_3}(t)$ and $e_{\alpha_3}(t)$. This also leads to $\|p_{31}(t)\|, \|p_{32}(t)\| > 0$, for all $t\geq0$ due to the selection of $\underline{b}_{r_3}, \bar{b}_{r_3}$ according to Section \ref{sec:bounds_select}.

\textit{Induction Step:} Now let us assume that the stability results of Theorem \ref{th:agent_k} hold for agents $3,\ldots,k-1$ (i.e., boundedness of all signals and satisfaction of \eqref{eq:e_h_bound} for all agents $3,\ldots,k-1$). Hence, one can verify that the results of \textit{Phase I} for agent $k$ still holds. Moreover, by employing the radially unbounded Lyapunov function candidate $V_k = (1/2) \sigma_k^T \sigma_k$, and since agent $k$ has its arbitrary two neighbors from the set $1,\ldots,k-1$, we can establish existence of an ultimate bound $\bar{\sigma}_k$ for $\sigma_k(t)$ in the same way as in \textit{Phase II}. Finally it is straightforward to repeat \textit{Phase III} and establish satisfaction of \eqref{eq:e_h_bound} for $e_{r_k}, e_{\alpha_{k}}$ along with $\|p_{ki}(t)\|, \|p_{kj}(t)\| > 0, (k,i), (k,j) \in \mathcal{E} \setminus \{(2,1)\}$ for all $t\geq0$, which finalizes the proof. 

\section{Derivation of $\dot{e}_{\alpha_k}$ in \eqref{eq: dot_alpha_k}}
\label{appen:alpha_tild_dot_calc}
\begin{figure}[tb]
	\centering
	%	\flushleft
	\begin{subfigure}[t]{0.24\textwidth}
		%		\centering
		\centering
		\includegraphics[width=\textwidth]{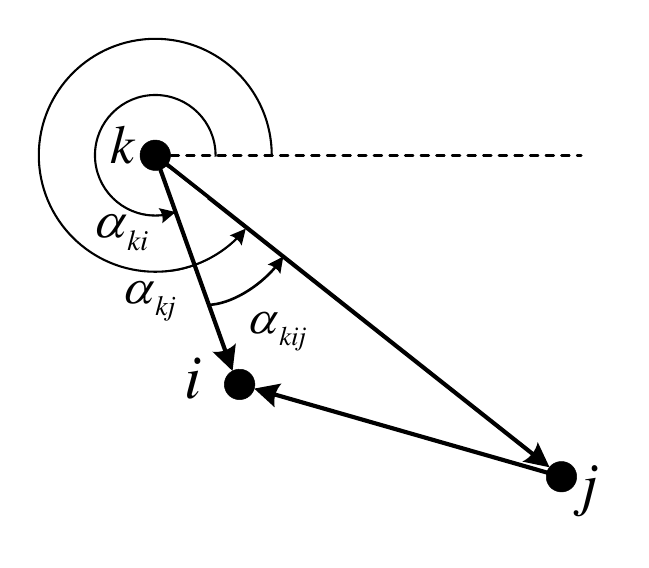}
		\caption{}
		\label{fig:trig_framework_alt}
	\end{subfigure}%\hspace{2cm}
	~
	\begin{subfigure}[t]{0.22\textwidth}
		%		\centering
		\centering
		\includegraphics[width=\textwidth]{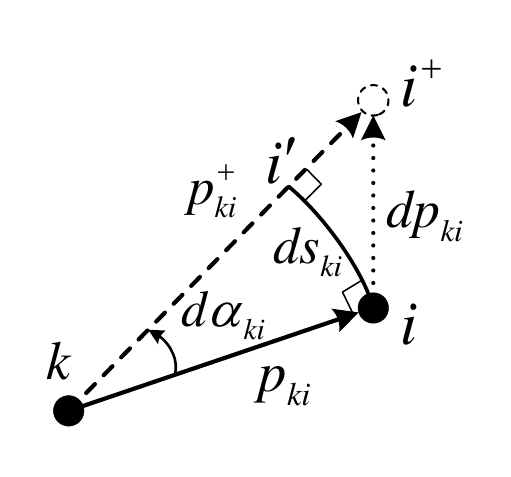}
		\caption{}
		\label{fig:infini_arc_angle}
	\end{subfigure}%\hspace{2cm}
	\caption{(a) Alternative edge-angle calculation. (b) Infinitesimal variation of  edge $(k,i)$'s angle.}
	\label{fig:sss}
\end{figure}

Consider a triangular sub-graph of $\mathcal{G}$, where $i<j<k$. An alternative way of calculating the edge-angle $\alpha_{kij}$ is given by \cite{liu2020distanceplus}:
\begin{equation} \label{eq:edge_angle_alt}
	\alpha_{kij} = \mathrm{mod} \lbrace \alpha_{kj} - \alpha_{ki}, 2\pi \rbrace, \quad (k,i), (k,j) \in \mathcal{E} \setminus \{(2,1)\},
\end{equation}
where $\alpha_{ki}$ and $\alpha_{kj}$ are the angles of the edges $(k,i)$ and $(k,j)$ measured counterclockwise from the x-axis of the global coordinate frame (see Fig.\ref{fig:trig_framework_alt}). Taking the time derivative of \eqref{eq:alpha_error} based on \eqref{eq:edge_angle_alt} yields:
\begin{equation} \label{eq:alpha_dot_alt}
	\dot{e}_{\alpha_k} = \dot{\alpha}_{kj} - \dot{\alpha}_{ki}.
\end{equation}
where $\dot{\alpha}_{ki}$ and $\dot{\alpha}_{kj}$ should be calculated explicitly. In this regard, consider ${p}_{ki}^{\,+} = {p}_{ki} + d {p}_{ki}$, where ${p}_{ki}^{\,+}$ represents the new relative position vector associated with edge $(k,i)$ subject to the infinitesimal changes in the positions of agents $k$ and $i$ that are captured by $d {p}_{ki}$. Notice that the infinitesimal motions of agents $k$ and $i$ modeled by $d {p}_{ki}$ can be seen as if only agent $i$ is moving. Therefore, for a better geometric representation, without loss of generality we assume that only agent $i$ has an infinitesimal motion as illustrated in Fig.\ref{fig:infini_arc_angle}. Moreover, assume that $d \alpha_{ki}$ represents the infinitesimal variation of $\alpha_{ki}$, and $d s_{ki}$ shows the infinitesimal variation of its corresponding curve with radius of $\|{p}_{ki}\|$. Since $\alpha_{ki}$ is in radians, from the arc length formula\footnote{It holds $ds = r d\theta$, where $r$ is the radii, $d\theta$ is the variation of the angle, and $ds$ is the variation of its corresponding arc length.} we get:
\begin{equation} \label{eq:arc_length_formul}
	d s_{ki} = \|{p}_{ki}\| d \alpha_{ki}.
\end{equation}
For the infinitesimal right triangle $\triangle_{ii^{'}i^{+}}$ we also have:
\begin{equation}\label{eq:arc_length_proj}
	d s_{ki} = (J z_{ki})^T d {p}_{ki},
\end{equation}
that is the projection of $d {p}_{ki}$ on the infinitesimal arc $d s_{ki}$. Invoking \eqref{eq:arc_length_formul} and \eqref{eq:arc_length_proj}, $\dot{\alpha}_{ki}$ is given by:
\begin{equation} \label{eq:alpha_ki_dot}
	\dot{\alpha}_{ki} = \dfrac{d \alpha_{ki}}{dt} = \dfrac{ z_{ki}^T}{\|{p}_{ki}\|} J^T \dot{{p}}_{ki}.
\end{equation}
A similar expression can also be obtained for $\dot{\alpha}_{kj}$. Therefore, using \eqref{eq:singledyn}, \eqref{eq:alpha_dot_alt}, and \eqref{eq:alpha_ki_dot} followed by the fact that $J^T = -J$, yields \eqref{eq: dot_alpha_k}.

\section{Quadratic form for $M_k \coloneqq G_k B_k \in \mathbb{R}^{2 \times 2}$} \label{appen:P.D. G_k B_k}
\begin{figure}[tb]
	\centering
	\includegraphics[scale=0.65]{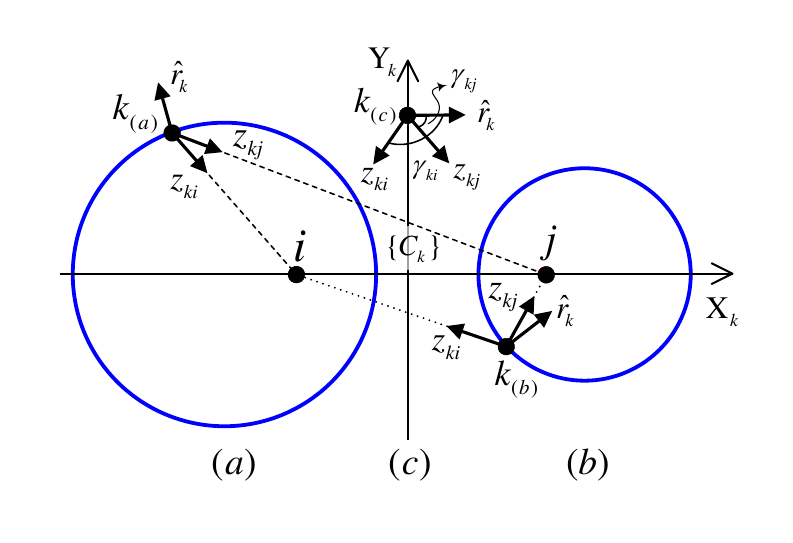}
	\caption{Configuration of $z_{ki}, z_{kj}$, and $\widehat{r}_k$ in three arbitrary positions of agent $k$ with respect to its neighbors.}
	\label{fig:PD_Pf}
\end{figure}

Consider matrices $B_k = [\widehat{r}_k \,|\, \widehat{\alpha}_k] \in \mathbb{R}^{2 \times 2} $ and $G_k \in \mathbb{R}^{2 \times 2}$, $k=3,\ldots,n$, as defined in the proof of Theorem \ref{th:agent_k}. Let $\eta_k \coloneqq (z_{kj} / \|p_{kj}\|) - (z_{ki} / \|p_{ki}\|) \in \mathbb{R}^{2}$ then $G_k = [\eta_k^T | \eta_k^T J]^T$.  Now $M_k \coloneqq G_k B_k$ gives:
\begin{equation}
	M_k =
	\begin{bmatrix}
		\eta_k^T \widehat{r}_k & \eta_k^T \widehat{\alpha}_k \\[3pt]
		\eta_k^T J \widehat{r}_k & \eta_k^T J \widehat{\alpha}_k 
	\end{bmatrix}
	=
	\begin{bmatrix}
		\eta_k^T \widehat{r}_k & - \eta_k^T J \widehat{r}_k \\[3pt]
		\eta_k^T J \widehat{r}_k & \eta_k^T \widehat{r}_k
	\end{bmatrix}
\end{equation}
where orthogonality of the bipolar basis is employed to obtain the right-hand side, that is: $\widehat{r}_k = J \widehat{\alpha}_k$ along with the fact that $J^{-1} = J^T = -J$. Let $x = [x_1 \; x_2]^T \in \mathbb{R}^{2}$, then one can verify that:
\begin{equation} \label{eq:quadratic_form}
	x^T M_k x = m_k (x_1^2 + x_2^2) = m_k \|x\|^2 
\end{equation}
where $m_k \coloneqq \eta_k^T \widehat{r}_k$. Since $\widehat{r}_k, z_{ki}, z_{kj}$ are unit vectors, from the (geometric) inner product formula we get:
\begin{equation}\label{eq:pd_proof_maineq}
	m_k = \eta_k^T \widehat{r}_k = \frac{\cos \gamma_{kj}}{\|p_{kj}\|} + \frac{-\cos \gamma_{ki}}{\|p_{ki}\|},
\end{equation}
where $\gamma_{ki}$ represents the (smallest) angle formed between $z_{ki}$ and $\widehat{r}_k$, and $\gamma_{kj}$ shows the (smallest) angle formed between $z_{kj}$ and $\widehat{r}_k$. In the sequel we will prove that $m_k>0$, which ensures positiveness of \eqref{eq:quadratic_form}. Consider three cases for agent $k$'s position with respect to its neighbors in the virtual Cartesian coordinate frame $\left\lbrace C_k \right\rbrace$ that are: (a) left half-plane, (b) right half-plane, and (c) on the $Y_k$ axis, as illustrated in Fig.\ref{fig:PD_Pf}. First, note that $\gamma_{kj} \leq \gamma_{ki}$ always holds. Without loss of generality, let us also assume that $\|p_{ki}\|, \|p_{kj}\|$ are bounded.

\textit{Case (a)}: Note that in this case $\widehat{r}_k$ is always directed outwards the $r_k = \mathrm{constant}$ curves. Moreover it always holds that $\pi/2 <\gamma_{ki}\leq \pi$, $0 \leq\gamma_{kj}\leq \pi$, and $\|p_{ki}\| < \|p_{kj}\|$. In this regard, whenever $0 \leq \gamma_{kj} \leq \pi/2$ then the first term in the right-hand side of \eqref{eq:pd_proof_maineq} is always positive or zero whereas the second term is always positive, thus $m_k > 0$ is ensured. Now consider when  $\pi/2 < \gamma_{kj} \leq \pi$. In this case the first term in the right-hand side of \eqref{eq:pd_proof_maineq} is always negative whereas the second term is always positive, however, due to $\gamma_{kj} \leq \gamma_{ki}$ and $\|p_{ki}\| < \|p_{kj}\|$ the second term always dominates the first term, thus we will always have $m_k > 0$. As a result we can conclude that for bounded $\|p_{ki}\|, \|p_{kj}\|$, we always have $m_k > 0$, whenever agent $k$ is in the left-half plane of $\left\lbrace C_k \right\rbrace$.   

\textit{Cases (b) and (c)}: Using similar arguments, we can also show that $m_k > 0$ whenever agent $k$ is on the $Y_k$ axis or in the right-half plane of $\left\lbrace C_k \right\rbrace$ for bounded $\|p_{ki}\|, \|p_{kj}\|$.   

Note that when $\|p_{ki}\|, \|p_{kj}\|$ become unbounded, $m_k$ may approach zero in all three cases (a), (b), and (c). However, unboundedness of $\|p_{ki}\|, \|p_{kj}\|$ is avoided in the proof of Theorem \ref{th:agent_k}. In particular, for $k=3$, from \textit{Phase I} of Theorem \ref{th:agent_k},  we have $\tilde{e}_3(t) \in \Omega_{\tilde{e}_3} = (-\underline{b}_{r_3}, \bar{b}_{r_3}) \times (-\underline{b}_{\alpha_3}, \bar{b}_{\alpha_3}), \forall t \in [0, \tau_{\mathrm{3,max}})$, hence followed by \eqref{eq:alpha_error}, \eqref{modu_err}, \eqref{eq:e_h_bound}, and Section \ref{sec:bounds_select} for the selection of $\underline{b}_{\alpha_3}, \bar{b}_{\alpha_3}$, we ensure that the edge-angle $\alpha_{312}$ is positively lower bounded away from zero and its upper bound is less than $2\pi$, $\forall t \in [0, \tau_{\mathrm{3,max}})$. Moreover, it is shown that $r_3$ and $\|p_{21}\|$ are bounded. These altogether are sufficient for boundedness of $\|p_{31}(t)\|, \|p_{32}(t)\|$, $\forall t \in [0, \tau_{\mathrm{3,max}})$. Note that similarly we can show that this property also holds for $k>3$, as explained in the induction step in the proof of Theorem \ref{th:agent_k}. Therefore, there always exists a positive lower bound $\underline{m}_k$ (depending on the choice of $\underline{b}_{\alpha_k}, \bar{b}_{\alpha_k}$) such that $0 <\underline{m}_k < m_k$.

%################################################################
\bibliographystyle{ieeetr}
\bibliography{Refs}

%###################################################
\end{document}